\documentclass[manuscript,screen]{acmart}

\AtBeginDocument{%
  \providecommand\BibTeX{{%
    \normalfont B\kern-0.5em{\scshape i\kern-0.25em b}\kern-0.8em\TeX}}}

\setcopyright{acmcopyright}
\copyrightyear{2023}
\acmYear{2023}
\acmDOI{XXXXXXX.XXXXXXX}

\acmConference[Conference acronym 'XX]{Make sure to enter the correct
  conference title from your rights confirmation emai}{June 03--05,
  2018}{Woodstock, NY}
\acmPrice{15.00}
\acmISBN{978-1-4503-XXXX-X/18/06}

\usepackage{xcolor}
\usepackage{hanging}
\usepackage{graphicx}
\usepackage{subcaption}
\usepackage{url}
\usepackage{enumitem}
\usepackage{tcolorbox}
\usepackage{float}
\usepackage{booktabs}
\usepackage{soul}
\usepackage[english]{babel} 
\usepackage{threeparttable}
\usepackage{tabularx}
\usepackage{longtable}
\usepackage{xltabular}
\usepackage{wasysym}
\usepackage{comment}
\usepackage{multicol, multirow}

\usepackage{todonotes}

\newcommand{\sectopic}[1]{\vspace{0em}\par\noindent{\textit{\bfseries #1}}}

\usepackage{varwidth}
\usepackage{ragged2e}
\usetikzlibrary{calc}

\newcounter{tipcounter}
\newcommand{\rev}[1]{\textcolor{black}{#1}}
\newcommand{\tosem}[1]{\textcolor{black}{#1}}

\begin{document}

\title{Replication in Requirements Engineering: the NLP for RE Case}

\author{Sallam Abualhaija}
\email{sallam.abualhaija@uni.lu}
\affiliation{%
  \institution{University of Luxembourg}
  \country{Luxembourg}
}

\author{Fatma Ba\c{s}ak Aydemir}
\affiliation{%
  \institution{Utrecht University}
  \country{The Netherlands}
  }
\email{f.b.aydemir@uu.nl}

\author{Fabiano Dalpiaz}
\affiliation{%
  \institution{Utrecht University}
  \country{The Netherlands}
}
\email{f.dalpiaz@uu.nl}

\author{Davide Dell'Anna}
\affiliation{%
  \institution{Utrecht University}
  \country{The Netherlands}
}
\email{d.dellanna@uu.nl}

\author{Alessio Ferrari}
\affiliation{%
 \institution{CNR-ISTI}
 \country{Italy}}
 \email{alessio.ferrari@isti.cnr.it}

\author{Xavier Franch}
\affiliation{%
  \institution{Technical University of Catalonia}
  \country{Spain}}
 \email{franch@essi.upc.edu}

\author{Davide Fucci}
\affiliation{%
  \institution{Blekinge Institute of Technology}
  \country{Sweden}}
\email{davide.fucci@bth.se}

\renewcommand{\shortauthors}{Abualhaija, et al.}

\begin{abstract}
 \textbf{[Context]} Natural language processing (NLP) techniques have been widely applied in the requirements engineering (RE) field to support tasks  such as classification and ambiguity detection. Despite its empirical vocation, RE research has given limited attention to replication of NLP for RE studies. 
Replication is hampered by several factors, including the context specificity of the studies, the heterogeneity of the tasks involving NLP, the tasks' inherent \textit{hairiness}, and, in turn, the heterogeneous reporting structure. \textbf{[Solution]} To address these issues, we propose a new artifact, referred to as \textsc{ID-Card}, whose goal is to provide a structured summary of research papers emphasizing replication-relevant information. We construct the \textsc{ID-Card} through a structured, iterative process based on design science. 
\textbf{[Results]} In this paper: (i) we report on hands-on experiences of replication, (ii) we review the state-of-the-art and extract replication-relevant information, (iii) we identify, through focus groups, challenges across two typical dimensions of replication: data annotation and tool reconstruction, and (iv) we present the concept and structure of the \textsc{ID-Card} to mitigate the identified challenges. 
\textbf{[Contribution]} This study aims to create awareness of replication in NLP for RE. We propose an \textsc{ID-Card} that is intended to foster study replication, but can also be used in other contexts, e.g., for educational purposes. 

\end{abstract}



\received{Date}
\received[revised]{Date}
\received[accepted]{Date}

\maketitle

\section{Introduction}
\label{sec:intro}

System requirements are primarily written in natural language (NL)~\cite{
fernandez2017naming, franch2023docum}. To analyze and manage textual requirements, the requirements engineering (RE) community has long been interested in applying natural language processing (NLP) technologies.  
The emerging research strand, namely NLP4RE, has received a lot of attention from both industry practitioners and academic researchers, leading to initiatives such as the NLP4RE workshop series~\cite{dalpiaz2018natural}.
To process textual information, NLP4RE tools utilize NLP techniques, which in turn rely on a variety of algorithms, dominated by machine learning (ML), deep learning (DL), and large language models (LLMs). A recent systematic mapping study~\cite{zhao2021natural} reports that the majority of  research papers in NLP4RE ($\approx$84\%) involve proposing novel solutions or validating existing technologies. However, only a small fraction of developed tools ($\approx$10\%) is made publicly available. Unavailable artifacts can impede the positioning of novel solutions through sound comparisons against existing ones and further hamper the industrial adoption of NLP4RE tools. 

\textit{Replication} is an important aspect of empirical evaluation that involves repeating an experiment under similar conditions using a different subject population~\cite{wohlin2012experimentation,shull2008role}. Replicability is currently regarded as a major quality attribute in software engineering (SE) research, \tosem{and it is one of the main pillars of Open Science~\cite{mendez2020open}}. The ACM badge system was introduced at the end of the 2010s\footnote{\url{https://www.acm.org/publications/policies/artifact-review-and-badging-current}. Visited April 5th, 2024} to reward,  among others, available and replicable research. Several major conferences---e.g., ICSE, ESEC/FSE, and RE---feature an Artifact Evaluation Track, which grants variants of the ACM badges. 
Two mapping studies, from Da Silva \textit{et al.}~\cite{da2014replication} and from Cruz \textit{et al.}~\cite{cruz2019replication}, cover replications in SE from 1994 to 2010, and from 2013 to 2018, respectively. In both studies, the RE field appears to be among the ones in which replication is most common. 
However, with a few exceptions (e.g., \cite{DBLP:conf/re/DalpiazDAC19}), replication does not seem commonly practiced in NLP4RE. This lack of replication can be attributed to various challenges, with one of the most prominent being the incomplete reporting of studies, as noted by Shepperd \textit{et al.}~\cite{shepperd2018role}.



In this paper, we propose a new artifact, referred to as \textsc{ID-card}, that fosters the replication of NLP4RE studies. The \textsc{ID-card} is a template composed of 47 questions concerning replication-relevant information, divided into seven topics. These topics characterize: the RE task addressed in the study; 
the NLP task(s) used to support the RE task; information about raw data, labeled datasets and  annotation process; implementation details; and information related to the evaluation of the proposed solution.
We advocate attaching the \textsc{ID-card}  as part of the submission for future NLP4RE papers as well as creating it in retrospect for existing papers. 
The \textsc{ID-card} can be created and/or used by authors, newcomers to the field, reviewers, and students. 

Defining the \textsc{ID-card} is triggered by our hands-on experience concerning two  replication scenarios. The first scenario involves replicating a state-of-the-art baseline~\cite{yang2011analysing} as part of building a benchmark for handling anaphoric ambiguity in textual requirements. The second scenario involves replicating a widely-used solution for classifying requirements into functional or non-functional~\cite{kurtanovic2017automatically} against which we compare and position a novel solution proposed by a subset of the authors of this paper, and previously published in~\cite{DBLP:conf/re/DalpiazDAC19}. Replication can be \textit{exact} when one follows the original procedure as closely as possible, or \textit{differentiated} when one adjusts the experimental procedures to fit the replication context~\cite{juristo2011role}. 
In our replication experience, exact replication was not possible mainly due to the incomplete reporting of implementation details in the original papers. Deciding on such unknown details during replication can alter the original study to some extent. Minimizing decisions due to unknown details is an essential motivation for  the \textsc{ID-card}. 
Building on our hands-on experience, we conducted several \textit{focus groups} through which we identified the different challenges that can be encountered with regard to extracting replication-relevant information from NLP4RE studies. To address these challenges, we defined an initial version of the \textsc{ID-card}. We then reviewed a representative sample from the NLP4RE literature and iteratively refined the \textsc{ID-card} according to our findings and observations.   

\begin{figure}
\centering
\includegraphics[width=0.9\textwidth] 
{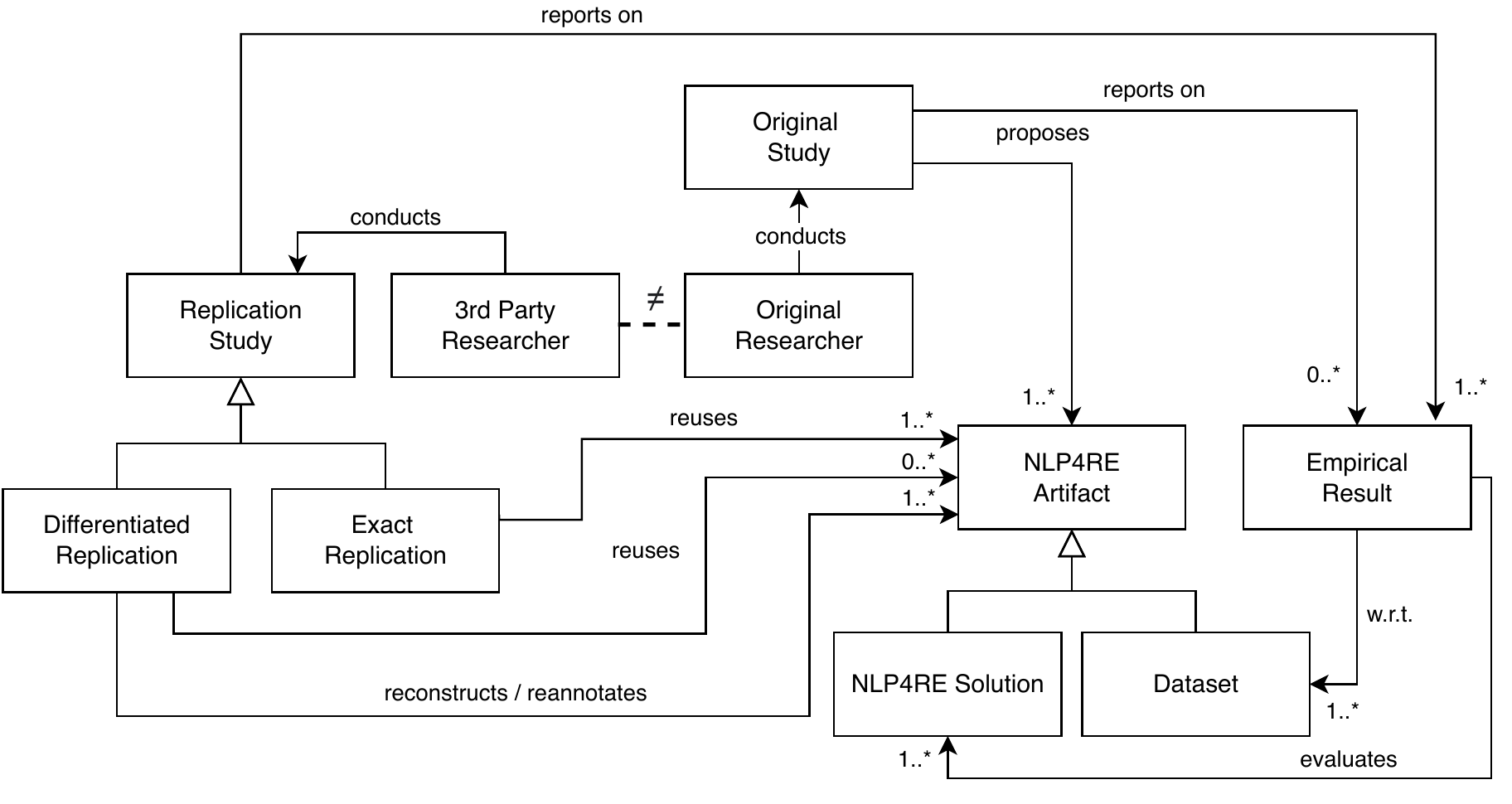}
\caption{Overview of the terminology used in this paper. }
\label{fig:terminology}
\end{figure}

\sectopic{Terminology.} Fig.~\ref{fig:terminology} provides an overview of the terminology used in this paper. 
\textit{Replication} refers to the attempt, conducted by \textit{third-party researchers} (different than the \textit{original researchers}), to obtain the same results of a specific \textit{original study}. \textit{Reuse} can be a prerequisite for replication: for replicating a study, third-party researchers may reuse the \textit{NLP4RE artifacts} (\textit{NLP4RE solutions} or \textit{datasets}) that were proposed and released by the original researchers. Note that the figure focuses on reuse in the context of replication, although researchers may also reuse artifacts for other purposes. 
We refer to the information that can be extracted from the original paper for the purpose of replication as \textit{replication-relevant information}. We further use \textit{solution} and \textit{tool} interchangeably, to indicate the automated procedure described in an NLP4RE paper to solve a particular problem. We use the term \textit{reconstruction} to denote the (re)-implementation of a solution as explained in the original paper. 

\sectopic{Contributions. } This paper makes the following contributions:
\begin{enumerate}
\item We identify a total of 16 challenges that can arise in practice during the replication of NLP4RE studies. We do so by relying on our hands-on experience where we have reconstructed two NLP4RE solutions in addition to performing an in-depth review of existing papers spanning diverse topics in the NLP4RE literature. As replication in NLP4RE often requires creating a dataset, we differentiate between the challenges related to dataset \textit{(re-)annotation}---aimed to determine the gold standard against which the solution is evaluated---versus challenges concerning tool \textit{reconstruction}. The final list of challenges is presented in Section~\ref{sec:challenges}. 
\item We devise an \textsc{ID-card} that summarizes in a structured way the replication-relevant information in NLP4RE papers. We demonstrate the applicability of the \textsc{ID-card} by manually creating the equivalent 
\textsc{ID-card}s for a total of 46 research papers from the NLP4RE literature. Specifically, we manually extracted the replication-relevant information from these papers and provided the answers to the questions posed in the \textsc{ID-card}. 
For 15 out of the 46 papers, we also let the original authors independently fill in the \textsc{ID-card}. \tosem{We also let the authors 
assess several aspects of the proposed \textsc{ID-card} such as its ease of use. 
}
As we discuss in Section~\ref{sec:id-card}, the results indicate that the \textsc{ID-card} can be indeed used as a complementary source to the original paper for facilitating replication. We make all the material publicly available in online annex~\cite{annex}.
\end{enumerate}

\sectopic{Structure.} \tosem{Section~\ref{sec:related} discusses related work about replicabilities and associated challenges.} Section~\ref{sec:rm} discusses the research method and research questions. Section~\ref{sec:cases} provides the context for the two replication scenarios considered in our work. Section~\ref{sec:challenges} describes the challenges of annotation and tool reconstruction.  Section~\ref{sec:id-card} presents the \textsc{ID-Card}. Section~\ref{sec:discussion} discusses issues related to the use of the \textsc{ID-Card} and the limitations of the study. Finally, Section~\ref{sec:conclusion} concludes the paper and sketches future directions.  

\section{Related Work}
\label{sec:related}

\tosem{
This paper focuses on replication in NLP4RE, a sub-field of RE that belongs to the broad area of SE. 
Only a few replication studies are published in this field---e.g., \cite{shah-safe-refsq-19,dabrowski-is-2023}, showing the need for methods and tools that can facilitate conducting such kind of studies.
Other studies in related domains have also identified some of the issues discussed in our paper~\cite{rahimi2022requirement,AAE22,zhao2023survey}, although those works build on secondary research (they are \textit{surveys}), while our approach heavily relies on our first-hand experience. 
In the following, we first discuss initiatives in Open Science and replicability in SE to show the relevance of this topic in the current discourse of the SE community and the complexity of ensuring replicability. Then, we refer to relevant works discussing guidelines and issues of study replication, to illustrate the addressed research gap.
}

\tosem{
\sectopic{Open Science and Replicability Initiatives.} Open Science in SE is a movement aiming to make research artifacts, including source code, datasets, scripts to analyze the data, and manuscripts, openly available to the research community~\cite{mendez2020open}. 
This enables \textit{replicability}---i.e., the possibility for other research teams to repeat a study by reusing the artifacts provided by the authors of the original work or by reconstructing them.
Moreover, Open Science facilitates verifiability and transparency, which are nowadays critical research drivers for SE in general, and RE in particular.
For example, ICSE 2024 review criteria define verifiability and transparency as ``The extent to which the paper includes sufficient information to understand how an innovation works; to understand how data was obtained, analyzed, and interpreted; and how the paper supports independent verification or \textit{replication} of the paper’s claimed contributions''\footnote{\url{https://conf.researchr.org/track/icse-2024/icse-2024-research-track\#Call-for-papers}. Visited January 4th, 2024}. 
Major SE outlets have introduced Open Science Policies (e.g., ESEC/FSE 2024\footnote{\url{https://2024.esec-fse.org/track/fse-2024-open-science-policy}. Visited January 4th, 2024}), requiring authors to make their artifacts available to the reviewers so that research results can be scrutinized during the review phase and made reproducible once published.
Moreover, conferences include Artifact Evaluation Tracks aimed at explicitly reviewing artifacts and assigning badges (e.g., the ones defined by ACM~\footnote{\url{https://www.acm.org/publications/policies/artifact-review-and-badging-current}. Visited January 4th, 2024}) to the associated papers.
These tracks typically include a separate Program Committee (PC) highlighting not only the relevance given to Open Science but also the \textit{complexity} of replicability assessment, which thus requires a dedicated PC.
The guidelines for artifact evaluation became quite extensive, as shown, for example, by the ICSE 2021 guidelines\footnote{\url{https://conf.researchr.org/getImage/icse-2021/orig/Submission.and.Reviewing.GuidelinesNUEVO.pdf}. Visited January 4th, 2024}, which count 13 pages, as the evaluation process needs to ensure the fulfillment of multiple fine-grained aspects.
To account for the complexity of replicability and the early degree of maturity of the community, the PC members are typically required to interact with the authors to provide them feedback to fix the artifacts before these can be evaluated---this is usually the case of software or scripts that cannot be properly run in order to produce the expected results.
In addition to Artifact Evaluation Tracks, the ROSE festival (Rewarding Open Science Replication and Reproduction in SE\footnote{\url{https://github.com/researchart/rose/tree/master}. Visited January 4th, 2024}) has been held in a number of SE top-venues. 
The festival includes enlightening talks about replication studies to raise awareness in the community.
In this scenario, not only conferences but also journals are also catching up with Open Science. The Empirical Software Engineering Journal (EMSE) introduced the OSI Open-data badge~\cite{mendez2019open} and ACM Transactions on Software Engineering Methodology (TOSEM) encourages Replicated Computational Results (RCR) Reports that should complement papers with information to support replications\footnote{\url{https://dl.acm.org/journal/tosem/replicated-computational-results}.  Visited January 4th, 2024}. This landscape of initiatives demonstrates the relevance given to replicability in SE, and the complexity of ensuring verifiability and transparency. }

\tosem{
\sectopic{Replicability Issues, Guidelines and Challenges.} 
From an online questionnaire of the \textit{Nature} journal involving 1,576 researchers from different disciplines, it emerged that 70\% of the respondents tried and failed to reproduce another scientist's experiment~\cite{baker20161}, indicating that replication is an issue for the wider research community. 
The main causes listed (60\% of the respondents) are pressure to publish and selective (i.e., incomplete) reporting. 
Furthermore, the absence of methods or code, as well as raw data, are regarded as relevant by more than 40\% of the participants. 
Among the solutions, the authors propose more academic rewards for replicability-enabling content, as well as journal checklists to ensure that replication-relevant material is included.   
}
\tosem{
Concerning the specific field of SE, multiple mapping studies and survey papers have been published on Open Science and replication, identifying the current status, guidelines, and open challenges.
Da Silva \textit{et al.}~\cite{da2014replication} cover replications in SE from 1994 to 2010.
The study highlights that replications have substantially increased in the period considered and discusses desired characteristics of the replication studies.  
The study also includes a useful quality checklist to evaluate replication studies based on Carver's guidelines~\cite{carver2010towards}. 
De Magalh{\~a}es \textit{et al.}~\cite{de2015investigations} covers replications from 1994 to 2013.
The authors lists recommendations oriented to those who perform replications, but also conditions that must be fulfilled by the original papers to be replicated. 
The latter include the need to provide all necessary replication-relevant details, precise and unambiguous definitions, and the sharing of raw data and tool versions. 
Gonzalez \textit{et al.}~\cite{gonzalez2012reproducibility,gonzalez2023revisiting} evaluates the replicability\footnote{The authors use the term \textit{reproducibility}, rather than replicability, but their definition of the term is similar to our definition of replicability: ``The measurement can be obtained with stated precision by a different team using the same measurement procedure, the same measuring system, under the same operating conditions, in the same or a different location on multiple trials. For computational experiments, this means that an independent group can obtain the same result using the author’s own artifacts''. For this reason, we use the term replicability to avoid confusion.} of Mining Software Repositories (MSR) studies, and propose a set of best practices to make a study replicable, applicable beyond the MSR field.
These are mainly focused on features of the replication package, which should include raw data, processed data, results, links to reused data (if any), and source code.
In the context of the NLP4RE, a secondary study by Ahmed et al.~\cite{AAE22} shows that published results in the area of requirements formalisation from natural language are often non-reproducible due to lack of access to tools, data, as well as critical information not reported in the primary studies.
Similarly, another systematic review in the same area~\cite{rahimi2022requirement} shows that the lack of openly accessible benchmarks hinders requirements formalisation research.}

\tosem{In a book chapter, Mendez \textit{et al.}~\cite{mendez2020open} characterize Open Science from multiple perspectives, including sharing manuscript preprints and artifacts, and provide detailed guidelines. 
The chapter is among the few works that discuss challenges related to Open Science. 
These include the overhead of sharing data, privacy, confidentiality, licensing, appropriate preparation of qualitative data, and anonymity issues---particularly in sharing pre-prints to be submitted to venues that adopt the double-blind review model.
These are mainly challenges from the perspective of the authors of the original study to be replicated, rather than from the third-party researchers who want to perform a replication.
Finally, Anchundia and Fonseca~\cite{anchundia2020resources} guidelines that can facilitate replications in SE and informally list challenges that prevent or hamper replications, namely ``reports and packages are neither sufficient to capture all information (e.g., raw material) nor to share tacit knowledge'', ``large amount of effort to obtain all necessary data'', ``replications do not satisfy professional needs'', `` the cost of conducting replications''. 
}

\tosem{
\sectopic{Research Gap.}
The landscape of Open Science initiatives in SE shows the increasing interest of the SE community in Open Science in general and replicability in particular underscoring that replication is a complex task, which requires further investigation and appropriate tool support.
Compared to previous studies, we notice that a large part of them list criteria to be fulfilled to enable replication~\cite{de2015investigations,gonzalez2012reproducibility,gonzalez2023revisiting,da2014replication}, but only two of them~\cite{anchundia2020resources,mendez2020open} identify  \textit{general} replication-related challenges.
None reports a \textit{context-specific} list of issues, specifically considering the viewpoint of third-party researchers who want to replicate a study, as in our case. While some general challenges also apply to our case, the provided list targets the NLP4RE context, which makes the identified issues more concrete.   
Furthermore, compared to previous work, our main outcome is a practical means to summarise relevant information to enable replications---i.e., the \textsc{ID-card}, which aims to address the identified challenges. 
}

\section{Research Questions and Method}
\label{sec:rm}
The research goal of this paper is to support the extraction and documentation of replication-relevant information from NLP4RE papers. 
%
To this end, we propose two main dimensions that characterize replication in the context of NLP4RE:
(i) the \textit{datasets} with their annotations, as research in NLP inherently relies on datasets for evaluation as well as for developing diverse solutions, e.g., training an ML classifier; and
(ii) the reconstruction of the proposed \textit{tools}, as most NLP4RE studies describe an automated NLP-based solution that tackles an RE problem~\cite{zhao2021natural}, which typically needs to be reconstructed. 

To achieve this goal, we devise a set of research questions (\textbf{RQs}). The first two questions are instrumental to respond to the third one:

\begin{enumerate}[start=1,label={\bfseries RQ\arabic*:}, leftmargin = 3em]
     \item \textit{What are the challenges of annotating datasets for training and evaluating NLP4RE tools?}\\
     As stressed in previous studies~\cite{zhao2021natural,dalpiaz2018natural,ferrari2017natural}, the number of \textit{annotated} (i.e., labeled) datasets in NLP4RE is scarce. Concrete annotation guidelines are essential to ensure the soundness of both reuse and replication. Reusing an existing (available) tool on different datasets 
     (e.g., based on industrial data) requires annotating these datasets
     following the same annotation procedure as the original dataset. 
     Replicating the entire study also requires applying the same annotation procedure.
     Missing guidelines can cause unwanted differences in results between the original study and the replicated one.
     In RQ1, we investigate the challenges of creating labeled datasets,  while making them available to and reusable by the research community. We also consider the case in which an existing dataset must be re-annotated for various reasons to support replication.  
    \item \textit{What are the challenges in reconstructing NLP4RE tools?}\\ 
    In RQ2, we study the challenges that third-party researchers 
    encounter when reusing and/or replicating existing tools as part of  their NLP4RE research. 
    Since the majority of NLP4RE tools are unavailable~\cite{zhao2021natural}, we highlight the process of \textit{reconstructing} (i.e., developing) a tool using only the descriptions of the algorithms provided in the research papers in which the tool is presented. More concretely, the notion of reconstruction in our context refers to the cases where the  solution is exclusively described in the research paper, yet is not complemented with any source code. While such cases can be regarded as extreme, they currently represent the common situation of tool reuse specifically in NLP4RE. Dealing with the extreme cases encapsulates as well the challenges that are pertinent to 
    less challenging replication scenarios. 
\end{enumerate}

\begin{enumerate}[start=3,label={\bfseries RQ\arabic*:}, leftmargin = 3em]
    \item \textit{How to support NLP4RE researchers to overcome the identified challenges in RQ1 and RQ2?}\\
    In response to RQ3, we propose  developing an artifact that complements the NLP4RE research papers to primarily facilitate replication. The artifact, which we refer to as \textsc{ID-card}, 
    is presented in Section \ref{sec:id-card}.
\end{enumerate}

\begin{figure}[!h]
\centering
\includegraphics[width=.8\textwidth]  {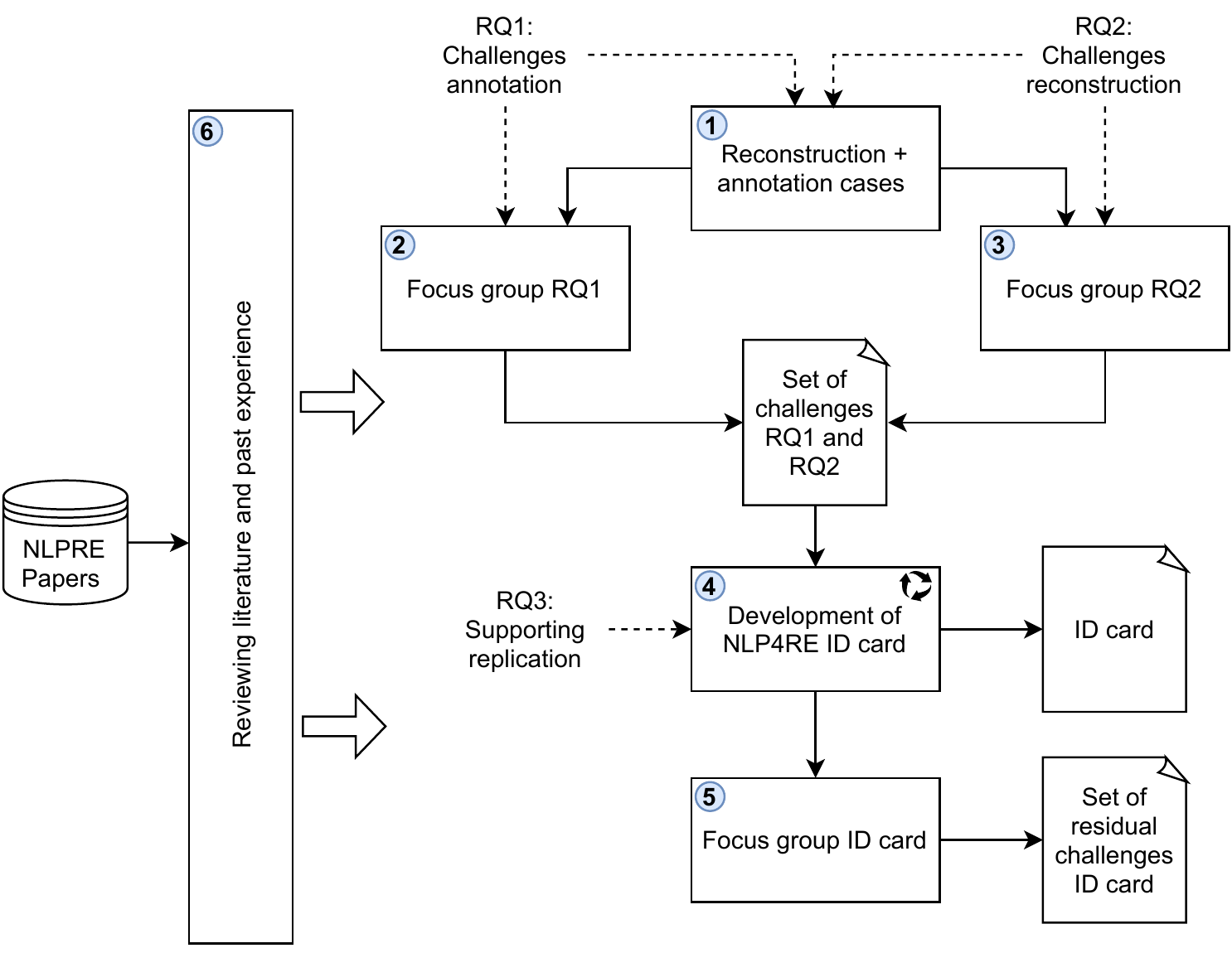}
\caption{\textcolor{black}{Overview of the research method of this paper. }}
\label{fig:r-method}
\end{figure}

To answer RQ1--RQ3, we follow the research method sketched in Fig.~\ref{fig:r-method}. The figure also serves as a reading guide for the various sections of this paper. To facilitate reading, the method is presented here linearly. However, the process was conducted iteratively. 

In order to respond to RQ1 and RQ2, we reconstructed two established NLP4RE tools for which the source codes were not available (see Step \textcircled{1} in Fig.~\ref{fig:r-method})~\cite{yang2011analysing,kurtanovic2017automatically}. The reconstruction naturally required the (re-)annotation of datasets. The first tool, originally proposed by Yang \textit{et al.} in 2011~\cite{yang2011analysing}, concerns anaphoric ambiguity handling, a long-standing, highly researched problem in NLP4RE~\cite{zhao2021natural}.  The second tool, originally proposed by Kurtanovic and Maalej in 2017~\cite{kurtanovic2017automatically}, concerns functional and non-functional requirements classification, a very popular RE task which has been also selected for the \textit{dataset challenge} at the IEEE Requirements Engineering  conference in 2017 (RE'17). These two tools were reconstructed by two subgroups of our research team at different times for different purposes, as we elaborate in Section~\ref{sec:cases}.

The rest of the process was mainly driven by focus groups, conducted according to the guidelines from Breen~\cite{breen2006practical}. Applying this research instrument, the cases and the collected data act as a trigger for reflection, so that the participants can brainstorm about specific challenges they encountered, recall previously faced problems, and compare their viewpoints with their peers.  We refer to Annex~\ref{sec:fg} for further details on how the focus groups were conducted (structure, participants, timing, analysis, etc.).

The  first two focus groups (Step \textcircled{2} and Step \textcircled{3} in Fig.~\ref{fig:r-method}) led to identifying a set of challenges related to RQ1 and RQ2. The challenges were reflected during the development of the NLP4RE \textsc{ID-card} (Step \textcircled{4}). 

Step \textcircled{4} included iterative activities consisting of multiple rounds of design and assessment among the authors. 
The activity was supported by an analysis of the literature (Step \textcircled{5} in Fig.~\ref{fig:r-method}) where we read and analyzed 46 representative NLP4RE papers, selected based on the literature review by Zhao \textit{et al.}~\cite{zhao2021natural}. \tosem{
A pair of researchers then used a similar strategy as the one applied in the mapping study to retrieve 155 additional papers published after 2019---i.e., the date of the mapping study.  
Finally, we selected 46 papers considering the following criteria: (1) Ensuring a balanced set of highly-cited papers (i.e., representing the state-of-the-art) as well as more recent papers. (2) 
Inspired by the topic categorization of the NLP4RE landscape~\cite{zhao2021natural}, our selected papers covered nine categories, including: requirements classification, tracing, defect detection, model generation, test generation, requirements retrieval, information extraction from both legal documents and requirements documents, and app review analysis. 
Specifically, we selected five papers from each category. 
}
At least two of the authors of this paper independently \tosem{manually analyzed 
} 
a subset of the research papers and then the findings were cross-checked (\textit{internal assessment}). The objective was to assess the completeness, cohesiveness, and coverage of the \textsc{ID-card}. 
In the final design iteration, we shared the NLP4RE \textsc{ID-card} with 15 authors of the NLP4RE papers which we considered in our design. Specifically, we let them  fill in the \textsc{ID-card} for their papers and further  evaluate their own experience by answering a questionnaire that we appended to the  \textsc{ID-card}, as we explain in Section~\ref{sec:eval} (\textit{external assessment}). 
To conclude, we organized a third focus group (Step \textcircled{6}) in which we discussed the \textsc{ID-card} and its use, leading to a number of observations to which we refer in the discussion (Section~\ref{sec:discussion}).

The research method of Fig.~\ref{fig:r-method} aligns with Wieringa's design science research, particularly with the design cycle~\cite{wieringa2014design} that comprises problem investigation, treatment design, and treatment validation.
In particular, \textit{problem investigation} is conducted in Steps \textcircled{1} through \textcircled{3}, wherein the researchers conduct reconstruction and annotation cases. These activities, complemented by the analysis of the literature conducted in Step \textcircled{5}, and consolidated by two focus groups, led to the identification of a set of challenges for RQ1 and RQ2. 
These challenges inform the \textit{treatment design}---the NLP4RE \textsc{ID-card} (Step \textcircled{4}). 
\textit{Treatment validation} of the \textsc{ID-card} has been conducted both via the internal and external assessment activities. 

In essence, multiple iterations of the design cycles were conducted as Step \textcircled{4} has a cyclic nature in Fig.~\ref{fig:r-method}. The last validation activity was conducted via a focus group (Step \textcircled{6}) in which the \textsc{ID-card} was finalized to its current form. 
\section{Reconstruction Cases of NLP4RE Solutions} \label{sec:cases}

We share our hands-on experience in replicating two state-of-the-art  solutions from the NLP4RE literature (Step\textcircled{1} in Fig.~\ref{fig:r-method}). The first solution (Sec.~\ref{subsec:ambiguity}) focuses on detecting anaphoric ambiguity, a specific type of defect in NL requirements, as presented in an early work by Yang et al.~\cite{yang2011analysing}. The second solution (Sec.~\ref{subsec:nfr}), presented in a more recent work by Kurtanovic and Maalej~\cite{kurtanovic2017automatically}, tackles the classification of non-functional requirements. Both defect detection and requirements classification are among the most common tasks in NLP4RE field ~\cite{zhao2021natural}. Furthermore, these topics and associated papers have garnered substantial recognition, with both papers accumulating over 160 citations according to Google Scholar\footnote{\tosem{Citations are collected on January 12, 2024.}}, making them highly representative of the NLP4RE field.
\subsection{Anaphoric Ambiguity Detection} \label{subsec:ambiguity} 

\sectopic{Motivation and overview.} 
Ambiguity in NL requirements is a long-standing research topic in RE~\cite{Kamsties:00,Berry:03,Kiyavitskaya:08,Dalpiaz:18,Gervasi:19,Ezzini:21}. Pronominal anaphoric ambiguity occurs when a pronoun can refer to multiple preceding noun phrases. 
The work of Yang \textit{et al.}~\cite{yang2011analysing} was chosen as a representative approach in anaphoric ambiguity detection. 
Since neither the dataset nor the source code was to be found publicly we decided to reconstruct the work using the details in the original paper and adapt it to the needs of our contest. This annotation activity acted as a trigger for the focus group in Step \textcircled{2}.  
\tosem{The RE literature discusses unconscious disambiguation where the stakeholder involved in a given project are able to disambiguate the requirements thanks to their domain knowledge~\cite{DeBruijnDekkers2010,PhilippoEtAl2013,RibeiroBerry2020,Ezzini:22}. Automated detection and resolution of different ambiguities is still a valid scenario in RE since not all stakeholders have the same level of domain knowledge and would hence resolve incorrectly. Another scenario is where end-to-end automation needs to be developed for other purposes, e.g., extracting information from requirements might require resolving referential ambiguity as a prerequisite.
} 


\sectopic{Annotation and Dataset Creation.}  
We used the PURE (PUblic REquirements) dataset~\cite{Ferrari:17} and randomly selected a subset of 200 requirement statements from seven domains such as railway and aerospace. \rev{Each requirement statement constituted one sentence, typically using the `shall' format. } The rationale behind our selection was that the resulting set should be manageable, time- and effort-wise, for annotation within the time frame of our contest. Further, covering diverse domains could be advantageous to assess the performance of a given solution across domains. 

Prior to annotating the requirements, we exchanged some guidelines about the ambiguity task. The task was to decide whether a pronoun occurrence in a given requirement is ambiguous or not by investigating the relevant antecedents. \rev{Each pronoun occurrence was analyzed by two annotators from the participants. }
Our annotation process was then performed in multiple rounds. After each round, we had a session to discuss our findings and some problematic issues.
The annotation process resulted in 103 ambiguous requirements (i.e., containing an ambiguous pronoun occurrence). For all remaining requirements which were marked as unambiguous, the annotators were asked to identify the antecedent they deemed correct. The outcome of our annotation process is a dataset where each pronoun
occurrence is labeled as ambiguous if it is  (i) marked as such by at least one annotator, or (ii) interpreted differently by the two annotators. Otherwise, when two annotators agree on the same interpretation, the pronoun occurrence is labeled unambiguous. \rev{We computed the pairwise inter-rater agreement using Cohen's Kappa~\cite{Landis:77} on a subset of $\approx$8\% from our dataset (16 requirements randomly selected). We obtained an average Kappa of 0.69, suggesting ``substantial agreement'' between the annotators. Following common practices, we used disagreements as indicators of ambiguity. }


\sectopic{Tool Reconstruction.} The solution proposed by Yang \textit{et al.}~\cite{yang2010methodology,yang2011analysing} combines ML and NLP technologies to detect anaphoric ambiguity in requirements, taking  a set of textual requirements as input and determining  whether each pronoun occurrence is ambiguous or not. We recently reconstructed this solution and publicly released it~\cite{Ezzini:22}.

\rev{The reconstructed version of the approach} includes four components: text preprocessing, pronoun-antecedent pairs identification, classification, and anaphoric ambiguity detection.
The first component (\textit{text preprocessing}) parses the textual content of the requirements document and applies an NLP pipeline consisting of four modules: (i) a tokenizer for separating out words from the running text, (ii) a sentence splitter for breaking up the text into separate sentences, (iii) a part-of-speech (POS) tagger for assigning a part of speech to each word in the text (e.g., noun, verb, and pronoun), and (iv) a chunker (or a constituency parser) to delineate phrases boundaries (e.g., noun phrases -- NPs).  
%

The second component (\textit{pronoun-antecedent pairs identification}) extracts all pronouns occurring from the input requirements, identifies a set of likely antecedents for each pronoun, and finally generates a set of pronoun-antecedent pairs. 
For the purpose of simplifying our reconstruction scenario and since coreference resolution was not directly relevant to our contest, we left it out. 

The third component (\textit{classification}) builds an ML-based classifier to classify a given pair of a pronoun $p$ and a relevant antecedent $a$ into \textit{YES} when $p$ refers to $a$, \textit{NO} when $p$ does not refer to $a$, or \textit{QUESTIONABLE} when it is unclear whether $p$ refers to $a$. The classifier is trained over a set of manually-crafted language features which characterize the relation between $p$ and $a$, e.g., whether $p$ and $a$ agree in number or gender. The original paper lists 17 features divided in three categories: 11 syntactic and semantic features, two document-based, and three corpus-based features.   
We dropped out \tosem{four} features that are computed using a proprietary library and related to sequential and structural information to facilitate the re-usability of our reconstructed solution. 

The last component (\textit{ambiguity detection}) applies a set of rules over the predictions produced by the ML-based classifier to distinguish the ambiguous cases. The original paper reports two thresholds for detecting correct and incorrect antecedents for unambiguous cases. 
In our \rev{reconstructed solution}, we redefined these thresholds with empirically optimized values for our dataset. The thresholds in the reconstructed tool could be generalized beyond the dataset as supported by empirical evidence in the paper, which reports the reconstruction~\cite{Ezzini:22}.
The reconstructed code is available in an online repository~\cite{ezzini:22-zenodo}.

\subsection{Functional and Non-functional Requirements Classification}
\label{subsec:nfr}

\sectopic{Motivation and Overview.}
Motivated by the importance of identifying quality aspects in a requirements specification starting from the early stages of software engineering, ML and NLP techniques have been widely used in proposing solutions to several related problems~\cite{montgomery2022empirical}, such as classifying requirements into functional vs. non-functional~\cite{cleland2007automated}\tosem{, which we refer to as the \textit{FR-NFR classification problem}}. Despite the variety of existing FR-NFR classification problem approaches, as of 2019, the most effective reported classifiers~\cite{kurtanovic2017automatically} relied on a characterization of the requirements at the word level, e.g., via text n-grams or Part-of-Speech (POS) n-grams, resulting in a large number of low-level features (100 or 500 in~\cite{kurtanovic2017automatically}). Moreover, the great majority of the literature in requirements classification focused, for evaluating the classifiers, on the PROMISE NFR data set~\cite{cleland2006detection}, a collection of 625 requirements from 15 projects created and classified by graduate students.

Aiming at a more interpretable solution than those relying on a large number of low-level (word-level) features, which make it hard for analysts to understand why the classifier performs well or poorly and why requirements are classified in a certain way, in 2019, \tosem{a subset of authors of this paper} proposed a new NLP approach to the FR-NFR classification problem~\cite{DBLP:conf/re/DalpiazDAC19}.
In proposing and evaluating the new solution, we (i) manually annotated a set of 1,500+ requirements consisting of 8 different projects, (ii) reconstructed Kurtanovic and Maalej's tool~\cite{kurtanovic2017automatically} to compare the proposed solution with the state-of-the-art.

\sectopic{Dataset and Annotation.}
We manually re-annotated 1,500$+$ requirements from the PROMISE dataset~\cite{cleland2006detection} and seven industrial projects. For the annotation, we followed an approach based on Li's taxonomy~\cite{li2014non}, which allows a requirement to possess both functional and non-functional aspects, as opposed to the original \tosem{annotation}, 
which allowed a single label per requirement. 
In particular, we annotated a requirement as possessing functional aspects (F) if it included either a
functional goal or a functional constraint, while we annotated a requirement as possessing quality aspects (Q) \tosem{(and thus being a non-functional requirement)} if it included a quality goal
or a quality constraint.
The decision on the functional aspect was independent of the
decision on the quality aspect; thus, a requirement could possess
only F aspects, only Q aspects, both aspects (F+Q), or none.
In the last case, we considered the requirement as denoting \tosem{auxiliary
} information~\cite{DBLP:conf/re/WinklerV16}.

\tosem{Each dataset was independently manually analyzed 
by two annotators. 
} Reconciliation meetings were then organized to review the
disputes in the annotation. If the annotators failed to convince each other,
a third annotator was consulted for the final label. The annotators went
over all disagreements and managed to resolve them.

\sectopic{Tool Reconstruction.}
To provide insights about how our approach compared against the state-of-the-art, we selected a relatively recent approach by Kurtanovic and Maalej~\cite{kurtanovic2017automatically}, described extensively in the original paper and shows excellent performance. 
The original solution consists of characterizing the training (PROMISE NFR) dataset using several low-level word features, such as n-grams, or POS n-grams. In particular, the authors of the original publication consider two cases with the top (most discriminant) 100 and 500 features. The requirements, labeled with the top features, are then used to train a Support Vector Machine (SVM) model with a linear kernel and use the trained model to classify previously unseen requirements as functional or non-functional.

Since the original classifier was not publicly available, 
we reconstructed it from the details provided in the original publication. We also complemented such information with a code of another classifier related to app reviews~\cite{kurtanovic2018user} that is partially available online and developed by the same \rev{research group}.

During reconstruction, we applied a few minor modifications to the original version (details in~\cite{DBLP:conf/re/DalpiazDAC19}). For example, to build parse trees, we used a different better-performing library (Berkeley’s \texttt{benepar}~\cite{DBLP:conf/acl/KleinK18}) than that used in the original publication (the Stanford parser~\cite{Chen:14}). We could not reproduce one of the classifier's features since the explanation in the original paper was
insufficient for us to make a correct re-implementation. We could not use the same  dataset  applied in the original solution to artificially balance the minority class of NFRs, as it was not publicly released. Finally, we released the reconstructed code in an online repository~\cite{fabiano_dalpiaz_2019_3309669}.

\section{Replication Challenges of NLP4RE Studies} \label{sec:challenges}

In this section, we present a total of 16 challenges which were derived through our first two focus groups complemented by our hands-on experience. Table~\ref{tab:challenges} lists the identified challenges in relation to dataset annotation (RQ1, Step~\textcircled{2} in Fig.~\ref{fig:r-method}) and tool reconstruction (RQ2, Step~\textcircled{3}).  

\begin{table}
  \caption{Summary of challenges pertinent to replication of NLP4RE studies. }\label{tab:challenges}
	\fontsize{8.05}{9.4}\selectfont
  \centering
  
    \begin{subtable}{\linewidth}
        \caption{Dataset Annotation Challenges (\textbf{RQ1})}
        \begin{tabularx}{0.98\textwidth} {@{} p{0.08\textwidth} @{\hskip 0.5em} X}
            \toprule
            \textbf{ID} & \textbf{Description} \\
            \midrule
            Ann1 & \rev{Some RE-specific categorization tasks lack solid theories that can guide the annotation process.} \\
            \midrule
            Ann2 & Besides annotation experience and theoretical knowledge, the lack of domain knowledge can limit the accuracy of the annotations. \\
            \midrule
            Ann3 & \rev{The annotation activity is time consuming due to factors such as language barriers, different individuals' background, and fatigue.} \\
            \midrule
            Ann4 & \rev{The annotation protocol can evolve and thereby necessitate the re-annotation of the data which might, again, cause additional time and effort.} \\
            \midrule
            Ann5 & \rev{Theoretical and practical training resources and opportunities are limited and not adequate for training novice annotators who are often trained during the annotation task by more experienced annotators. }\\
            \midrule
            Ann6 & \rev{The lack of benchmarks entails that annotated datasets enabling comparison against the state-of-the-art are scarce.} \\
            \midrule
            Ann7 & \rev{Available imbalanced datasets pose the challenge of both understanding the minority class and consequently the annotation of new examples thereof.} \\
            \midrule
            Ann8 & \rev{Determining the right amount of context to be shared alongside the annotation raw data is essential and can significantly affect the annotation results.} \\
            \midrule
            Ann9 & \rev{Motivating the annotators poses another challenge since an immediate observation of the impact of a given annotation task is not always possible. } \\
            \midrule
            Ann10 & \rev{Annotators are often not experienced in managing the social aspects or resolving conflicts originating from power, authority, or other social relations.} \\
            \bottomrule
        \end{tabularx}
      \label{tab:table_a}
    \end{subtable}

    \begin{subtable}{\linewidth}
	  \vspace*{1em}
	  \caption{Tool Reconstruction Challenges (\textbf{RQ2})}
	  \begin{tabularx}{0.98\textwidth} {@{} p{0.08\textwidth} @{\hskip 0.5em}  X}
    	    \toprule
    	    \textbf{ID} & \textbf{Description} \\
    	    \midrule
            Rec1 & \rev{The reconstruction-relevant information and implementation details of the original approach can be ambiguous, imprecise, and incomplete.} \\
            \midrule
            Rec2 & \rev{If a tool was partially or fully developed and/or evaluated using proprietary data, then there is no guarantee that the reconstructed tool would be identical to the original one since the used data cannot be accessed for reconstruction purposes.} \\
            \midrule
            Rec3 & Communication with the original authors is not always useful since the actual information sources may not be available anymore.\\
            \midrule
            Rec4 & \rev{The continuous evolution of the NLP ecosystem entails that some libraries become outdated, unavailable, or not maintained anymore.} \\
            \midrule
            \textcolor{black}{Rec5} & \tosem{Tools are typically developed as prototypes and not maintained in the long term. } 
            \\
            \midrule
            \textcolor{black}{Rec6} & \tosem{Tool reconstruction is not (yet) valued as a self-standing research contribution, and hence researchers are discouraged to replicate tools overtime. } 
            \\
    	    \bottomrule
	  \end{tabularx}
	  \label{tab:table_b}
	\end{subtable}   

\end{table}

\subsection{Challenges of Dataset Annotation and Re-annotation (\textbf{RQ1})}
\label{subsec:rq1}

During our focus group for RQ1, we identified 10 challenges pertinent to creating a labeled dataset for a specific RE task. We organize the challenges into four categories: (a) \textit{theoretical foundation} \rev{(Ann1 and Ann2), (b) \textit{annotation process} (Ann3 -- Ann5), (c) \textit{dataset-related} (Ann6 -- Ann8), and (d) \textit{human and social aspects-related} (Ann9 and Ann10).}

\subsubsection{Theoretical Foundation Challenges} 

%
\paragraph{\textbf{Ann1. Some RE-specific categorization tasks lack solid theories that can guide the annotation process.}} 
Unlike typical NLP tasks, which rely on solid theories from linguistics, RE categorization tasks often lack a common and agreed-upon theory defining the different classes. 
For example, thus far, researchers have proposed numerous definitions and taxonomies for classifying requirements into functional or non-functional~\cite{glinz2007non,li2014non}. 
For instance, the requirement \textit{``The system shall backup the customer data every 2 hours.''} can be labeled as non-functional according to Glinz's definition \cite{glinz2007non}, since it conveys a specific quality concern (i.e., reliability). Conversely, the exact requirement can be labeled as both functional and non-functional according to Li \textit{et al.}'s definition~\cite{li2014non}, since it has a functional part (i.e., to backup customer data) as well as a quality that qualifies it in terms of performance (i.e., every 2 hours).
Since neither of these two definitions prevails over the other, the annotation output can differ depending on the annotators' background. %



\paragraph{\textbf{Ann2. Besides annotation experience and theoretical knowledge, the lack of domain knowledge can limit the accuracy of the annotations.}}
For example, the lack of knowledge about regulations, standards, and data types when annotating a dataset can result in confusing compatibility with compliance requirements. Furthermore, the annotators may often use existing datasets from a particular domain (e.g., automotive) without necessarily having adequate experience in that domain or any interactions with domain experts or the original annotators who created the datasets. For example, not knowing whether an abbreviation such as STANAG 4609 refers to a standard or regulation leads to confusion for identifying compatibility and compliance requirements, respectively. 

\subsubsection{Annotation Process Challenges} 
\paragraph{\textbf{Ann3. The annotation activity is time consuming due to factors such as language barriers, different individuals' background, and fatigue.}}
An annotation task in NLP4RE often deals with manually analyzing several requirements artifacts and assigning labels as required by the respective task. Additional delays can be caused by several factors, including but not limited to: 
(i) \textit{Language barriers:} Most requirements texts are written in English. \tosem{Since English is not necessarily the mother tongue of those who write the requirements or those who read them, 
} handling requirements in English may lead to grammatical errors in the text and/or different interpretations. 
(ii) \textit{Reconciliation of multiple annotations for the same sample:} Individuals' background  might be a barrier to reaching full agreement. even when the different annotators see each others' perspectives, sometimes they still insist on their own. 
(iii) \textit{Fatigue:} The amount of effort spent on the annotation task causes considerable fatigue which can affect the performance of the annotators. Defining clear fatigue thresholds and fatigue-mitigation strategies require finding a balance between throughput and quality of the resulting annotations. 

\paragraph{\textbf{Ann4. The annotation protocol can evolve and thereby necessitate the re-annotation of the data which might, again, cause additional time and effort.}}
To increase inter-rater agreement, the annotation activity is often conducted iteratively, holding regular consolidation meetings. This may result in changes in the annotation protocol, which may compromise the quality of the annotation. \rev{Any change would immediately raise the question about} whether to re-annotate the previous annotations and thus improve the quality or continue without doing so due to time considerations.

\paragraph{\textbf{Ann5. Theoretical and practical training resources and opportunities are limited and not adequate for training novice annotators who are often trained during the annotation task by more experienced annotators.}}
In an ideal scenario, annotators should be trained prior the annotation task, because (i) annotators are expected to meet a minimum  qualification level (e.g., domain knowledge, motivation, availability, seriousness, etc.), and (ii) annotators will be more attentive when they learn about certain issues, e.g., the interrater agreement. However, 
appropriate level of details and training resources are often difficult to determine in advance. 
An alternative is to prepare guidelines and/or have small iterations (pilots to train/test people). Written guidelines, despite being useful, can be ambiguous (in particular as future references after the annotation has been conducted). Hence, the same results are unlikely to be obtained by different annotators even if the same written guidelines are provided. 

\subsubsection{Dataset-related Challenges} 
\paragraph{\textbf{Ann6. The lack of benchmarks 
entails that annotated datasets enabling comparison against the state-of-the-art are scarce. }}
For example, currently the most widely adopted dataset for the FR-NFR classification problem is the PROMISE dataset. This is a benchmark with questionable quality, since it contains requirements which were written and classified by students. More recent work in the NLP4RE literature has provided a re-annotated version of that dataset~\cite{li2014non,DBLP:conf/re/DalpiazDAC19}, clearly showing some discrepancy with the original annotated dataset.  

\paragraph{\textbf{Ann7. Available imbalanced datasets pose the challenge of both understanding the minority class and consequently the annotation of new examples thereof.}}
The different labels are often unevenly distributed among the data. Given that there are few examples that are associated with the minority label, annotators will have less examples to learn about that label.  For instance, the distributions of functional and non-functional requirements in our replication study were quite diverse and often skewed across the different annotated datasets leaving the annotators with doubts when it comes to the minority class. 

\paragraph{\textbf{Ann8. Determining the right amount of context to be shared with the raw data to be annotated is essential and can significantly affect the annotation results.}}
We observed in both reconstruction cases that not knowing the wider context of a given requirement to be annotated can complicate the annotation process. In the case of referential ambiguity, the context of a requirement may clarify the ambiguity. Similarly, in the case of non-functional requirements classification, the location of the requirement in the document or the surrounding requirements may hint the intended class. 

\subsubsection{Human and Social Aspects-related Challenges} 
\paragraph{\textbf{Ann9. Motivating the annotators poses another challenge since an immediate observation of the impact of a given annotation task is not always possible.}}
Inviting annotators from industry is difficult and sometimes not possible. Annotation per se has no direct application (or interest) in industry since practitioners (domain experts in the case of annotation) do not need annotated data, but rather a working solution. In addition, merely annotation tasks are not always appreciated in the research community and are thus hard to publish. The latter observation is confirmed by the majority of the focus group participants, who 
regard annotation as an unfulfilling task.  


\paragraph{\textbf{Ann10. Annotators are often not experienced in managing the social aspects or resolving conflicts from power, authority, or other social relations.}}
The human aspects of an annotation process (e.g., the role of persuasion) are typically not covered by the annotation protocols and training. However, the social aspect is often necessary both to improve the understanding of each others' perspectives and consequently reaching agreements. For any annotation task, social challenges cover different aspects including the decision of objectively accounting for persuasion, the best schema to decide on consensus (e.g., consensus seeking, majority votes, senior one makes the final decision), motivating the annotators, power and authority relations, and the impact of peer-pressure.

\subsection{Challenges of Tool Reconstruction (\textbf{RQ2})} \label{subsec:rq2}

During our focus group for RQ2, we identified a total of five challenges pertinent to tool reconstruction
concerning (a) \textit{information availability} \rev{(Rec1 -- Rec3)}, (b) \textit{technological divergence} \tosem{(Rec4 and Rec5)}, 
and (c) \textit{motivation} \tosem{(Rec6)}. 

\subsubsection{Information Availability Challenges} 


\paragraph{\textbf{Rec1. The reconstruction-relevant information and implementation details of the original approach can be ambiguous, imprecise, and incomplete.}}
The original paper is normally written to communicate a study rather than with the goal of facilitating tool reconstruction. Therefore, details about the architecture are frequently left out. While omitting implementation details can make a paper more readable, missing such details can complicate tool reconstruction. Examples of such implementation details include parameters' choice, seed values for the experiments, configuration options, detailed features descriptions, and (the exact versions of) libraries. 
The availability of the source code is not always useful for reconstruction since the accompanying documentation is often unclear. 
%
For instance, one of the major obstacles we encountered  in reconstructing the requirements classification tool was related to understanding the  learning features for the ML classifier. The authors \rev{of the original tool} applied 11 different \textit{types} of learning features, including text n-grams, POS n-grams, the fractions of nouns, verbs, adjectives, adverbs in the requirement, etc. 
While some of these features are standard and can be easily re-implemented (e.g., the depth of the syntax tree), other features were unclear, and their descriptions in the original paper were inadequate for an exact 
re-implementation. An example of unclear feature concerns the unigrams of part-of-speech (POS) tags on the clause and phrase level. 

\paragraph{\textbf{Rec2. If a tool was partially or fully developed and/or evaluated using proprietary data, then there is no guarantee that the reconstructed tool would be identical to the original one since the used data cannot be accessed for reconstruction purposes.}}
Private or unpublished datasets can be sometimes used in developing (some steps of) the original tool \rev{or during validation. For example, 
in the considered FR-NFR classification problem, the original tool applied, in addition to the public PROMISE dataset, another dataset from Amazon software reviews to balance the minority class.} 
This second dataset was also used for selecting the most relevant learning features. However, as we did not have access to this additional dataset, we had to resort to a different solution for balancing the minority class, compromising thereby the selection of the same relevant features.
Since the evaluation set also contained proprietary data, verifying the correctness of the reconstructed tool was in this case not possible. 


\paragraph{\textbf{Rec3. Communication with the original authors is not always useful since the actual information sources may not be available anymore.}}
\rev{It is fundamental to communicate with the authors of the original tool for assistance with the reconstruction. To this end, in some cases limited responsiveness can be observed, in particular when the authors are contacted about very specific details (e.g., explanation of learning features). Ideally, the authors of the original tool should help validate the correctness of the reconstructed tool. }
However, in practice, the developers of a tool are often PhD or MSc students, who may have left the organization, taking away in-depth knowledge about the tool. 

\subsubsection{Technological Divergence Challenges.} 

\paragraph{\textbf{Rec4. The continuous evolution of the NLP ecosystem entails that some libraries become outdated, unavailable, or not maintained anymore.}}
An existing approach from the NLP4RE research often contains text preprocessing steps based on available NLP technologies (e.g., stopwords removal or lemmatization). Outdated NLP libraries do not always have equivalent replacements. 
Based on our experience, older libraries were applied, for example, in the case of requirements classification to construct syntax trees or undersample the majority class. 
Some of these old libraries became deprecated, incompatible with new versions of libraries doing other tasks, or performed worse than their respective new versions. Updating libraries can therefore  introduce a discrepancy between
what is expected and what is achieved, for the better or worse. For instance, to  create the parse trees accurately in one of our reconstructions, we opted for Berkley’s \texttt{benepar} library instead of Stanford parser used in the original paper. 
This challenge was more notable during the reconstruction of the  ambiguity handling tool which was proposed in 2011. 
Since then, the NLP technologies have been through several breakthroughs, changing thereby the landscape for textual preprocessing. 

\paragraph{\textbf{Rec5. Tools are typically developed as prototypes and not maintained in the long term. }}
\tosem{Developed NLP4RE tools are often left in their initial prototype status and not maintained in the long term. As a consequence, tools become harder to reuse over time: the availability of platforms and operating systems changes over time and specific supporting software (which itself may no longer be available) can block reconstruction. One of the reasons emerging from our focus groups for lack of maintenance is the focus on novelty in NLP4RE research where tools are used predominantly for publication purposes.}

\subsubsection{Motivation-related Challenges}
\paragraph{\tosem{\textbf{Rec6. Tool reconstruction is not (yet) valued as a self-standing research contribution, and hence researchers are discouraged to replicate tools overtime.} 
}}
Similar to the annotation task,  reconstructing a tool is not sufficiently rewarding from a community standpoint. 
While some venues accept replications in the form of scientific evaluation paper types which can involve tool reconstruction, such papers could still be harder to publish. The reason for that according to the focus group's participants is due to the RE community rating novelty aspects higher than 
 consolidation of knowledge. 
\section{Design and Evaluation of the \textsc{ID-Card} (RQ3)}\label{sec:id-card}

Once the challenges related to data annotation and tool reconstruction were identified through the responses to RQ1 and RQ2, we conducted Step~\textcircled{4} of our research method and implemented the concept of \textsc{ID-card}. 

\subsection{\textsc{ID-Card} Design}

\rev{We designed the structure of the \textsc{ID-card} following} an iterative method composed of three steps: 1) Preliminary Definition; 2) Internal Assessment; and 3) External Assessment. 

\sectopic{Preliminary Definition.} In the first step, we outlined a list of information items needed for reconstructing an NLP4RE solution based on our own experience in reconstructing the two tools introduced in Section~\ref{sec:cases}, as well as on the challenges gathered while answering RQ1 and RQ2.
The researchers worked in pairs, and each pair drafted a set of questions and possible answers for a specific dimension. The considered dimensions included: task, dataset, annotators \& annotation process, tool, and evaluation. The dimensions were selected based on previous experience and brainstorming between the researchers. After the pairs of researchers drafted the questions, \tosem{we carried out a group meeting for one hour to consolidate them. } At this stage, the researchers listed 56 questions. Another two iterations were carried out during the Internal Assessment (step 2, which is described below) to reach a stable set of 47 questions that are simplified to be more 
understandable by possible target readers. 

\sectopic{Internal Assessment.} We analyzed 46 papers from the NLP4RE literature (five of which co-authored by at least one of the authors of this paper), and extracted replication-relevant data according to the information items from the first step. 
We included in our analysis also the two papers which we used in our replication scenarios (see Section~\ref{sec:cases}). 
%

Following the collection of the 46 selected papers, we used the preliminary \textsc{ID-card} (56 closed questions) where at least one researcher 
extracted information from each of the 46 papers. \tosem{For four papers, we also independently filled in the \textsc{ID-card} for the same paper to have some common manual analysis 
in the same topic category which we cross-checked and discussed at a later stage. }
The goal of this activity was to assess the applicability of the \textsc{ID-card} on a broad set of topics and possibly customizing it for different categories. After that, the researchers had a plenary discussion to share the problems emerged during this initial application of the \textsc{ID-card} and decide how to mitigate these problems. 
The outcome of this meeting was an updated list of 47 questions in total, including a combination of  32 closed and 15 open questions.

\sectopic{External Assessment.} The \textsc{ID-card} was shared with the authors of the original papers. Specifically, we asked them to fill in the \textsc{ID-card} for their papers without sharing our previously filled-in cards. 
We contacted 32 authors, avoiding to contact the same research group for multiple papers, and received filled-in \textsc{ID-card}s from 15 authors for 15 different papers. 
For each paper, we analyzed the discrepancies between our \tosem{filling } 
and the \tosem{filling } 
of the original authors and further provided explicit notes listing the points of disagreement and reflections on possible motivations. 
In a plenary discussion of 1.5 hours, we discussed the identified discrepancies based on the researchers' notes to identify major sources of disagreement. \tosem{Most of the disagreements were due to the level of details provided. Different answers were provided with different levels of granularity to some questions in the \textsc{ID-card}. For example, in the case of multiple datasets used in the same study, some answers focused only on one dataset whereas others provided information about all datasets. Other disagreements were observed in the NLP task that is used in the study. In this case, we acknowledge that multiple NLP technologies can be applied in the same paper for solving some RE task. }
\tosem{The goal of our plenary session was to identify the elements that could lead to misunderstanding, gather external viewpoints on the developed card, and discuss possible residual issues. 
After this analysis, we rephrased some of the questions according to our observations to make them more concrete and reduce possible disagreements. }

\subsection{\textsc{ID-Card} Description}
The \textsc{ID-card} resulting from the various iterations consists of 47 questions, divided into seven sections, described below.   A compact version of the \textsc{ID-card} is displayed in Table~\ref{tab:idcard}. \rev{The card contains questions that either attempt to fully or partially address the challenges introduced in Section~\ref{sec:challenges} or aim to provide metadata about the research paper such as questions~I.1~and~II.1. }
We provide in an online annex~\cite{annex}: the complete \textsc{ID-card}, validation material, and two \tosem{filled-in 
\textsc{ID-card}s of the reconstructed tools described in Section~\ref{sec:cases}. } 

\begin{description}
\item[\textbf{I. RE Task.}] This section identifies the RE task addressed in the paper, 
e.g., classification, tracing, defect detection. The \textsc{ID-card} provides several options from which only one option can be selected. The options include the nine categories listed earlier and an additional option that enables 
adding a unanticipated RE task. 
While the majority of NLP4RE papers address one main RE task, this is not always the case. For example, a paper could describe multiple RE tasks such as generating a model from requirements specifications and a completeness checking method to identify incompleteness in requirements according to the generated model. Each RE task can be solved using a combination of NLP tasks such as text classification, named entity recognition, and semantic role labeling. Some papers also introduce different datasets and  develop multiple tools. In this case, the \textsc{ID-card} is intended to be filled in separately for each of
the RE tasks. The rationale behind this decision is two-fold. First, we simplify the overall design of the \textsc{ID-card} (e.g., we avoid the reiteration of certain questions for each RE task). Second, by decomposing the work into distinct RE tasks, we facilitate the retrieval of replication-relevant information associated with the paper, which is the main goal of the \textsc{ID-card}. Even when the entire work in such a paper is considered for replication, decomposing the work into distinct RE tasks helps better understand and replicate the tools more accurately.   

\item[\textbf{II. NLP Task(s).}] This section specifies the NLP tasks used to support the RE task, i.e., classification, translation, information extraction and information retrieval (again, unanticipated NLP tasks may be specified). NLP tasks are distinguished from RE tasks as some NLP tasks could be applied for different purposes. For instance, classification (NLP task) can be used to organize requirements into different categories, but also to detect defects or identify trace links (RE tasks). More than one NLP task can be selected since the NLP tasks can be used in combination---e.g., in model generation, one uses information extraction \textit{plus} translation. 
\item[\textbf{III. NLP Task Details.}] This section characterizes the details of the RE task addressed, i.e., the RE task input granularity (e.g., document, word, paragraph, etc.), and the output type. We define various  options of the output types that differ for each NLP task. For example, a classification task requires  specifying whether the output is binary-single label (e.g., ambiguous XOR not ambiguous), multi-class multi-label (e.g., feature request OR bug OR praise), or other options in between. One has to further specify the possible labels (i.e., classes) of the output. For translation, the output can be text, but also test cases or models. Specific to translation, one has to specify the \textit{cardinality}, i.e., one input to one output (e.g., one document to one diagram) many input to many output (e.g., from many use case descriptions to one class diagram and multiple other diagrams). 
\item[\textbf{IV. Data and Dataset.}] This section characterizes the dataset used in the paper. In particular, one has to provide details about: the size of the dataset; the year; the raw data source data (e.g., proprietary industrial data, regulatory documents, user generated content---eight options plus ``other''); the level of abstraction of the data (e.g., user-level, business-level, system-level). In this section, the term \textit{data} collectively identifies both possible input and output data of the previously selected NLP task(s). Therefore the questions support multiple answers. This choice was driven by the need to keep the \textsc{ID-card} well-structured and easy to retrieve.   
This section also asks for information concerning the format of the data (use case, ``shall'' requirements, diagrams, etc.), the degree of rigor (unconstrained NL, restricted grammar, etc.), and the actual language (if applicable). Additional questions are included about the heterogeneity of the dataset---in terms of domain coverage and number of sources from where the data is obtained---as well as about the data licensing and the URL to access the dataset. 
\item[\textbf{V. Annotators and Annotation Process.}] This section includes information to characterize the annotators, in terms of background knowledge, number, and level of bias, in case annotation was carried out on the raw data. In addition, information is collected about the  adopted annotation scheme (if any) and the process to measure and resolve disagreements.  \tosem{We note that this section focuses mainly on manual annotation, a common practice for creating datasets in NLP4RE. With the current NLP technologies, researchers are shifting towards creating datasets using automated means. Extending the annotation section to cover the replication of automated annotation is left for future work.  }
\item[\textbf{VI. Tool.}] This section collects information about any implementation provided alongside the paper---e.g., scripts, executable programs, APIs, collectively designated with the term \textit{tool}. Specifically, the questions in this section require information about the enabling technology of the implemented NLP solution  (e.g., machine learning, rule-based), what has been released (e.g., binary file, source code, etc.), and additional information concerning documentation, licensing, dependencies and other details that can help access and execute the tool. 
\item[\textbf{VII. Evaluation.}] This section requires information about the evaluation carried out in the paper. Specifically, the evaluation metrics (precision/recall, AUC, etc.), the type of validation process (cross-validation, train-test split, etc.), and the baselines used for comparison (if applicable). \tosem{Investigating other evaluation alternatives, e.g., the impact of using the tool on the downstream development, is left for future work. }
\end{description}

\tosem{We note that the \textsc{ID-card} provides a comprehensive view of what information is relevant for replication. However, for a particular paper, one might fill in only some sections of the \textsc{ID-card} which are found in and relevant to the paper. For example, if the paper only presents a new dataset without an automated approach, then only the section about annotation might be relevant. } 

\subsection{\tosem{Tracing Challenges to Questions in the \textsc{ID-card}.  }}  
In the following, we discuss the rationale for the majority of the questions in the \textsc{ID-card}, in relation with the challenges described in Table~\ref{tab:challenges}. \tosem{We note that the traceability is not one-to-one mapping between challenges and the \textsc{ID-card}. Some challenges such as the little value given by our RE community to replicating studies cannot be handled by the \textsc{ID-card}.   }  

Recall that the classification of the RE tasks \tosem{(Question I.1)} was adopted from the recent systematic literature review by Zhao \textit{et al.}~\cite{zhao2021natural}, while the NLP task classification \tosem{(Question II.1) was specifically designed to complement the RE tasks}. 
%
%
%
The \textsc{ID-card} partially addresses the annotation challenges (see Table~\ref{tab:challenges}) as follows: 
\begin{itemize}
\item To address challenge Ann1, questions V.5 and V.6 are introduced to highlight the necessity for establishing a clear annotation protocol prior to the annotation process and further making it publicly available afterwards. Question V.10 also points out that communicating among annotators is required for achieving a high-quality dataset. Once the annotation protocol is agreed on, the likelihood that it frequently evolves (Ann4) is limited.

\item While including domain experts in the annotation process is not always possible, questions V.1 -- V.4 in the \textsc{ID-card} inform the researchers interested in reconstructing a dataset about whether the domain expertise was involved or not. 

\item With regard to Ann3, question V.8 indicates the need to mitigate fatigue. 

\item Questions IV.10 -- IV.13 are concerned with the dataset. The questions spotlight the publicly available datasets alongside the licenses under which they are released. By leveraging such datasets in future similar annotation tasks, one can address the challenges Ann5 -- Ann7.  

\item Question V.7 gives insights about the context that is shared with the annotators during the annotation process. The \textsc{ID-card} can be used to create a common practice to assist researchers in designing new annotation tasks. Thus, addressing the challenge Ann8. 
\end{itemize}

With regard to the tool-reconstruction challenges, the \textsc{ID-card} demands for more details and precise information concerning the implementation of the tool which are often omitted from a paper due to space limitations. Specifically, questions VI.6 and VI.8 are about the libraries employed in the implementation, question VI.5 is about the available documentation related to the tool, and questions VI.3 and VI.9 are about what has been publicly released.

\begin{table}
\footnotesize
\caption{\textsc{ID-card} for NLP4RE research papers. } %
\label{tab:idcard}
\centering
\begin{tabularx}{\textwidth} {@{} p{0.07\textwidth} | @{\hskip 0.5em}  X}
\toprule
\multicolumn{2}{c}{\textbf{I. RE Task}}   \\
\cmidrule(lr){1-2}
\textbf{I.1.} & \textbf{What RE Task is your study addressing?} $\Circle$~Requirements Classification $\Circle$~Requirements tracing $\Circle$~Requirements defect detection $\Circle$~Model generation $\Circle$~Test generation $\Circle$~Requirements retrieval $\Circle$~Information extraction from legal documents $\Circle$ ~App review analysis $\Circle$~Dependency and relation extraction $\Circle$~Information extraction from requirements $\Circle$~Other $\ldots$\\
\midrule
\multicolumn{2}{c}{\textbf{II. NLP Task(s)}}   \\
\cmidrule(lr){1-2}
\textbf{II.1.} & \textbf{What types of NLP task is your study tackling?} $\Square$~Classification (choose among classes) $\Square$~Translation (models, tests, etc.) \tosem{$\Square$~Information Extraction (glossary, terms) } $\Square$~Information Retrieval (search / rank) $\Square$~Other $\ldots$\\
\midrule
\multicolumn{2}{c}{\textbf{III. NLP Task Details}}   \\
\cmidrule(lr){1-2}
\textbf{III.1.} & \textbf{What is the input of your NLP task?} $\Square$~Document $\Square$~Paragraphs $\Square$~Sentences $\Square$~Phrases $\Square$~Words $\Square$~Structured/tabular text $\Square$~Other $\ldots$\\
\textbf{III.2.} & \textit{[Only for Classification]} \textbf{What type of classification is the study about?} $\Square$~Binary-Single label
$\Square$~Binary-Multi label
$\Square$~Multi class-Single label $\Square$~Multi class-Multi label $\Square$~Other $\ldots$\\
\textbf{III.3.} & \textit{[Only for Classification]} \textbf{What are the labels that can be assigned?} $\ldots$ \\
\textbf{III.4.} & \textit{[Only for Information Extraction]} \textbf{What is the level of granularity of the extracted elements?} $\Square$~Sentences $\Square$~Phrases $\Square$~Words $\Square$~Other $\ldots$\\
\textbf{III.5.} & \textit{[Only for Information Extraction]} \textbf{What is the type of the extracted elements?} $\ldots$ \\
\textbf{III.6.} & \textit{[Only for Translation]} \textbf{What is the type of output?}  $\Square$~Text $\Square$~Table $\Square$~Graphical diagram $\Square$~Executable model $\Square$~Test cases $\Square$~Other $\ldots$\\
\textbf{III.7.} & \textit{[Only for Translation]} \textbf{What is the translation mapping cardinality between initial input and final output?} $\Square$~1~to~1 $\Square$~1~to~many $\Square$~Many~to~1 $\Square$~Many~to~many $\Square$~Other $\ldots$ \\ 
\textbf{III.8.} & \textit{[Only for Information Retrieval]} \textbf{What are the types of elements retrieved with the query?} $\Square$~Document $\Square$~Paragraphs $\Square$~Sentences $\Square$~Phrases $\Square$~Words \\
\textbf{III.9.} & \textit{[Only for Other NLP Task]} \textbf{What is the type of output?} $\ldots$ \\
\midrule
\multicolumn{2}{c}{\textbf{IV. Data and Dataset}}   \\
\cmidrule(lr){1-2}
\textbf{IV.1.} & \textbf{How many data items do you process?} $\ldots$ \\
\textbf{IV.2.} & \textbf{In which year or interval of years was the data produced?}  $\ldots$ \\
\textbf{IV.3.} & \textbf{What is the source of the data?} $\Square$~Industrial project, proprietary data
$\Square$~Industrial project, publicly available data
$\Square$~Community-based open source projects $\Square$~Textbook examples or cases $\Square$~Student projects $\Square$~Toy requirements $\Square$~Legal/regulatory documents $\Square$~User generated content $\Square$~Other $\ldots$ \\ 
\textbf{IV.4.} & \textbf{What is the level of abstraction of the data (not limited to requirements)?} $\Square$~User-level $\Square$~Normative-level $\Square$~Business-level $\Square$~System-level $\Square$~Module-level $\Square$~Code-level $\Square$~Other $\ldots$ \\
\textbf{IV.5.} & \textbf{What is the format of the data?} $\Square$~``Shall''-requirements $\Square$~User stories $\Square$~Use cases $\Square$~User reviews $\Square$~Social media posts $\Square$~Bug/defect reports $\Square$~Messages in user forums \tosem{$\Square$~Other $\ldots$ }\\
\textbf{IV.6.} & \textbf{How rigorous is the format of the data?} $\Square$~Unconstrained natural language (NL) $\Square$~Template-based controlled NL $\Square$~Restricted grammar based controlled NL $\Square$~Semantically-augmented NL $\Square$~Other $\ldots$ \\
\textbf{IV.7.} & \textbf{What is the natural language of the data (if applicable)?} $\ldots$ \\
\textbf{IV.8.} & \textbf{Please list which domains your data belongs to:} \textit{[comma separated]} $\ldots$ \\
\textbf{IV.9.} & \textbf{From how many different sources does your data come from?} $\ldots$ \\
\textbf{IV.10.} & \textbf{Is the dataset publicly available (also from other authors)?}  $\Circle$~Fully $\Circle$~Partially  $\Circle$~Upon request $\Circle$~No $\Circle$~Other $\ldots$ \\
\textbf{IV.11.} & \textbf{What license has been used?} $\Circle$~No license $\Circle$~Reuse only for non-commercial purposes $\Circle$~Reuse for any purpose $\Circle$~Modification only for non-commercial purposes $\Circle$~Modification for any purpose $\Circle$~Other $\ldots$ \\
\textbf{IV.12.} & \textbf{Where is the dataset stored?} $\Circle$~On a private/corporate website $\Circle$~In a repository like GitHub $\Circle$~In a persistent platform with DOI $\Circle$~Other $\ldots$ \\
\textbf{IV.13.} & \textbf{Provide a URL to the dataset, if available, or to the original paper that proposed the dataset}  $\ldots$ \\
\midrule 
\multicolumn{2}{c}{\textbf{V. Annotators and Annotation}}   \\
\cmidrule(lr){1-2}
\textbf{V.1.} & \textbf{How many annotators have been involved?} $\ldots$ \\
\textbf{V.2.} & \textbf{How are the entries annotated?} $\Circle$~Multiple annotator per entry $\Circle$~One annotator per entry (no quality control) $\Circle$~One annotator per entry (quality control) $\Circle$~Partly multiple annotators, partly one annotator per entry $\Circle$~Other $\ldots$ \\
\textbf{V.3.} & \textbf{What is the average level of application domain experience of the annotators?}  $\Circle$~None or unknown  $\Circle$~Informed Outsider  $\Circle$~Domain expert  $\Circle$~Other $\ldots$ \\
\textbf{V.4.} & \textbf{Who are the annotators?} $\Circle$~The designer of the technique/tool $\Circle$~People who have direct contact with designers $\Circle$~Independent annotators (e.g., crowd workers) $\Circle$~Other $\ldots$ \\
\textbf{V.5.} & \textbf{How was the annotation scheme established among the annotators?} $\Circle$~Only via labels $\Circle$~Oral agreement among the annotators $\Circle$~Written guidelines with label definitions $\Circle$~Written guidelines with definitions and examples $\Circle$~Other $\ldots$ \\
\textbf{V.6.} & \textbf{Did the authors make the written guidelines public?} $\Circle$~YES $\Circle$~NO \\
\textbf{V.7.} & \textbf{Did the authors share other information that could support the annotators other than the elements to annotate?} $\Circle$~NO $\Circle$~Surrounding context $\Circle$~Entire document $\Circle$~Other $\ldots$ \\
\textbf{V.8.} & \textbf{Did the authors employ techniques to mitigate fatigue effects during the annotation sessions?} $\Circle$~YES $\Circle$~NO \\
\textbf{V.9.} & \textbf{What are the metrics used to measure intercoder reliability?} $\Circle$~Cohen's Kappa $\Circle$~Krippendorf's Alpha $\Circle$~Fleiss' Kappa $\Circle$~Other $\ldots$ \\
\textbf{V.10.} & \textbf{How were conflicts resolved?} $\Square$~Majority voting $\Square$~Discussion among annotators $\Square$~Resolution by authors $\Square$~Resolution by independent expert (not an annotator) $\Square$~Disagreements were disregarded $\Square$~Not resolved $\Square$~Other $\ldots$ \\
\textbf{V.11.} & \textbf{What is the measured agreement?} $\ldots$ \\ 
\bottomrule
\end{tabularx}
\end{table}

\begin{table}
\ContinuedFloat
\footnotesize
{\color{black}\caption{\textsc{ID-card} for NLP4RE research papers (continued).} %
\label{tab:idcard}
\centering
\begin{tabularx}{\textwidth} {@{} p{0.07\textwidth} | @{\hskip 0.5em}  X}
\toprule
\multicolumn{2}{c}{\textbf{VI. Tool}}   \\
\cmidrule(lr){1-2}
\textbf{VI.1.} & \textbf{What is the type of proposed solution?} $\Square$~Supervised machine learning (ML) $\Square$~Unsupervised ML $\Square$~Supervised deep learning (DL) $\Square$~Unsupervised DL $\Square$~Rule-based $\Square$~Other $\ldots$ \\ 
\textbf{VI.2.} & \textbf{What algorithms are used in the tool?} $\ldots$ \\ 
\textbf{VI.3.} & \textbf{What has been released?}  $\Square$~Standalone Tool  $\Square$~Service on the web  $\Square$~Library/API  $\Square$~Source code  $\Square$~Executable notebook  $\Square$~Pre-trained models  $\Square$~No tool has been released  $\Square$~Other $\ldots$ \\ 
\textbf{VI.4.} & \textbf{What needs to be done for running the tool?} $\Circle$~No installation is needed $\Circle$~Import and integrate into your own code $\Circle$~Compile and run $\Circle$~Virtual machine/Docker container $\Circle$~Reproduce the tool from the explanation of the paper $\Circle$~Other $\ldots$ \\ 
\textbf{VI.5.} & \textbf{What type of documentation has been provided alongside the tool?} $\Square$~README file $\Square$~Pseudocode/illustration in the paper $\Square$~Wiki or dedicated website $\Square$~Tutorial $\Square$~Ready-to-use examples $\Square$~An academic paper $\Square$~No documentation $\Square$~Other $\ldots$ \\ 
\textbf{VI.6.} & \textbf{What type of dependencies does the tool have?} $\Square$~None $\Square$~Open-source libraries/software $\Square$~Proprietary libraries/software $\Square$~Specific operating system $\Square$~Specific hardware $\Square$~External knowledge-bases $\Square$~Other $\ldots$ \\ \textbf{VI.7.} & \textbf{How is the tool released?} $\Circle$~On a private/corporate website $\Circle$~In a repository like GitHub $\Circle$~In a persistent platform with DOI $\Circle$~Other $\ldots$ \\ 
\textbf{VI.8.} & \textbf{What license has been used? Other} $\Circle$~No license $\Circle$~Reuse only for non-commercial purposes $\Circle$~Reuse for any purpose $\Circle$~Modification only for non-commercial purposes $\Circle$~Modification for any purpose $\Circle$~Other $\ldots$ \\
\textbf{VI.9.} & \textbf{Where is the tool released?} \textit{[url]} $\ldots$ \\ 
\midrule
\multicolumn{2}{c}{\textbf{VII. Evaluation}}   \\
\cmidrule(lr){1-2}
\textbf{VII.1.} & \textbf{What metrics are used to evaluate the approach(es)?} $\Square$~Precision/Recall $\Square$~F-score $\Square$~Accuracy $\Square$~AUC $\Square$~MAP $\Square$~LAG $\Square$~Word Error Rate (WER) $\Square$~NIST -- METEOR -- ROUGE -- BLEU $\Square$~Other $\ldots$ \\ 
\textbf{VII.2.} & \textbf{What is the validation procedure?} $\Circle$~Train-test split $\Circle$~Cross-validation $\Circle$~Cross-project validation $\Circle$~Entire dataset $\Circle$~Other $\ldots$ \\ 
\textbf{VII.3.} & \textbf{What baseline do you compare against?} 
$\Square$~Existing tool or algorithm $\Square$~Reconstructed tool from other research papers $\Square$~Automated, but self-defined $\Square$~Theoretical/Conceptual $\Square$~Human baseline $\Square$~None $\Square$~Other $\ldots$ \\ 
\textbf{VII.4.} & \textbf{Please provide more details about the baseline you compare against, if any.} $\ldots$ \\ 
\bottomrule
\end{tabularx}
}\end{table}



\subsection{Evaluation of the ID Card} 
\label{sec:eval}
\begin{figure}
\centering
\begin{subfigure}{\textwidth}
  \centering
  \includegraphics[width=\textwidth]
  {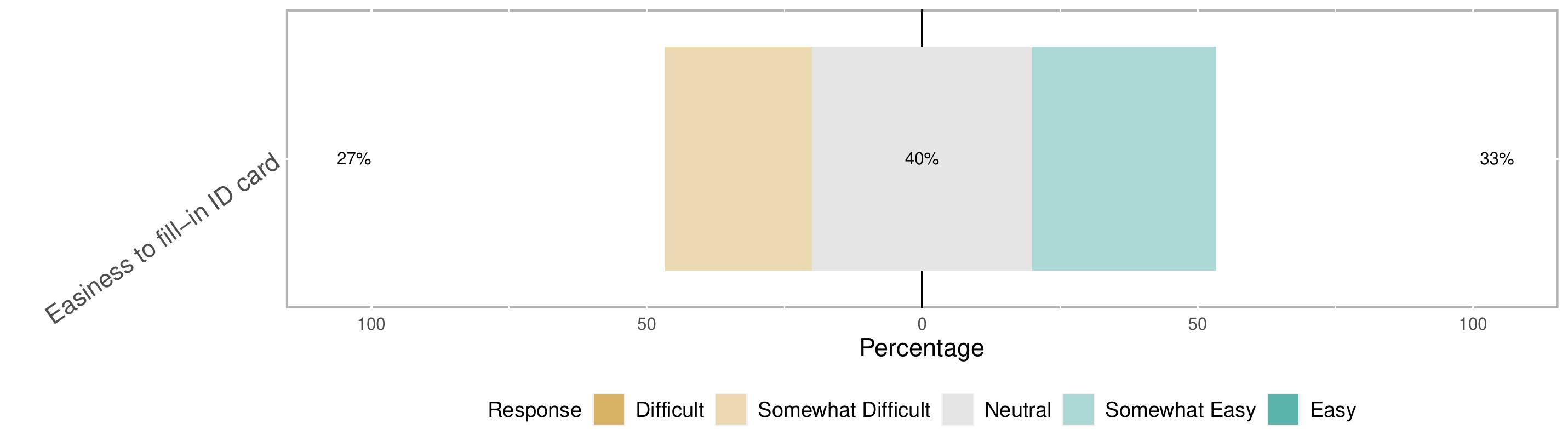}
  \caption{\textsc{ID-card} perceived ease of use.}
  \label{fig:use}
\end{subfigure}
\\

\begin{subfigure}{\textwidth}
  \centering
  \includegraphics[width=\textwidth]
  {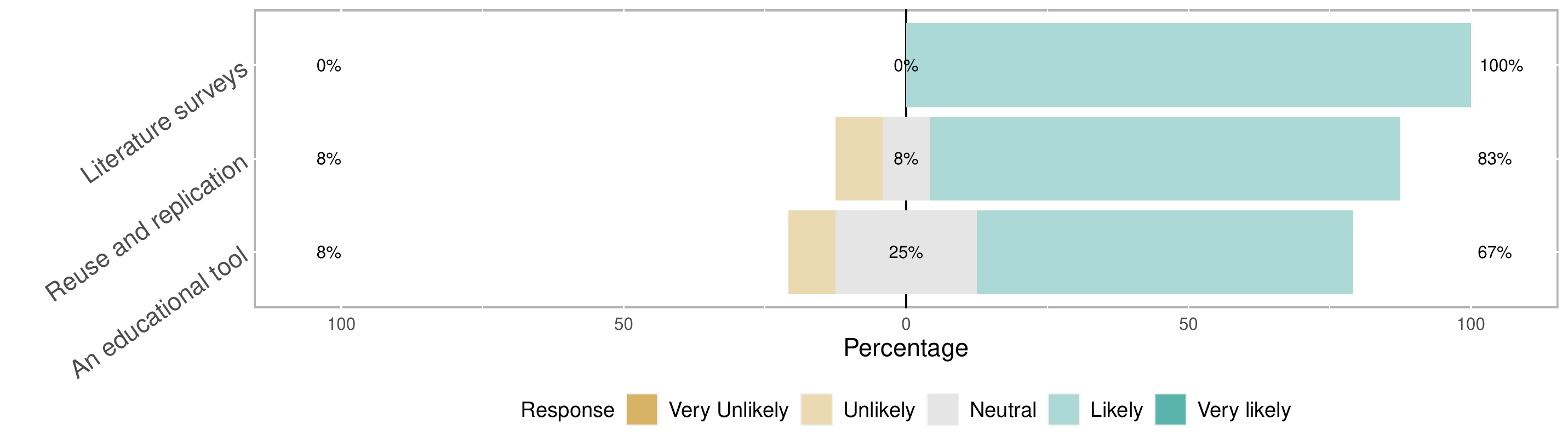}
  \caption{\textsc{ID-card} intention to use for different use cases.}
  \label{fig:intention}
\end{subfigure}
\caption{Results of TAM assessment for the NLP4RE \textsc{ID-card}.}
\label{fig:tam}
\end{figure}

We disseminated a short survey (four Likert-scale questions and two open-ended questions) among the 15 authors who filled in the \textsc{ID-card}, asking them to provide explicit feedback. 
Inspired by the questionnaire from the Technology Acceptance Model~\cite{Dav89}, we included survey items (see Fig.~\ref{fig:tam}) regarding the perceived ease of use and intended use \rev{in three different use cases, reuse and replication, literature surveys (as typical instrument that NLP4RE newcomers use for learning the state of the art), and education}.
The vast majority of respondents agreed that the \textsc{ID-card} has the potential to be used \rev{for the purposes stated above}. 
Nevertheless, opinions were mixed regarding the card ease of use. 
In the open feedback section of the survey, some respondents indicated a preference for the support of multiple datasets and a more interactive format which support conditional sections based on previous answers. 

\begin{figure}[!h]
\centering
\begin{subfigure}{\textwidth}
  \centering
  \includegraphics[width=\textwidth]
  {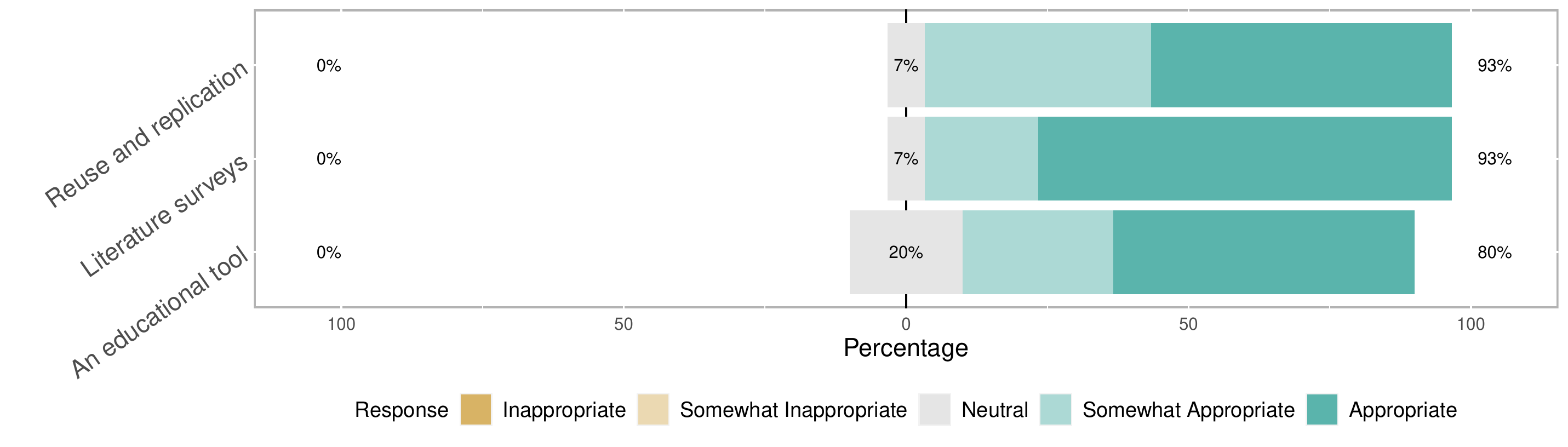}
  \caption{NLP4RE ID Card perceived appropriateness for different use cases.}
  \label{fig:appropriateness}
\end{subfigure}
\\

\begin{subfigure}{\textwidth}
  \centering
  \includegraphics[width=\textwidth]
  {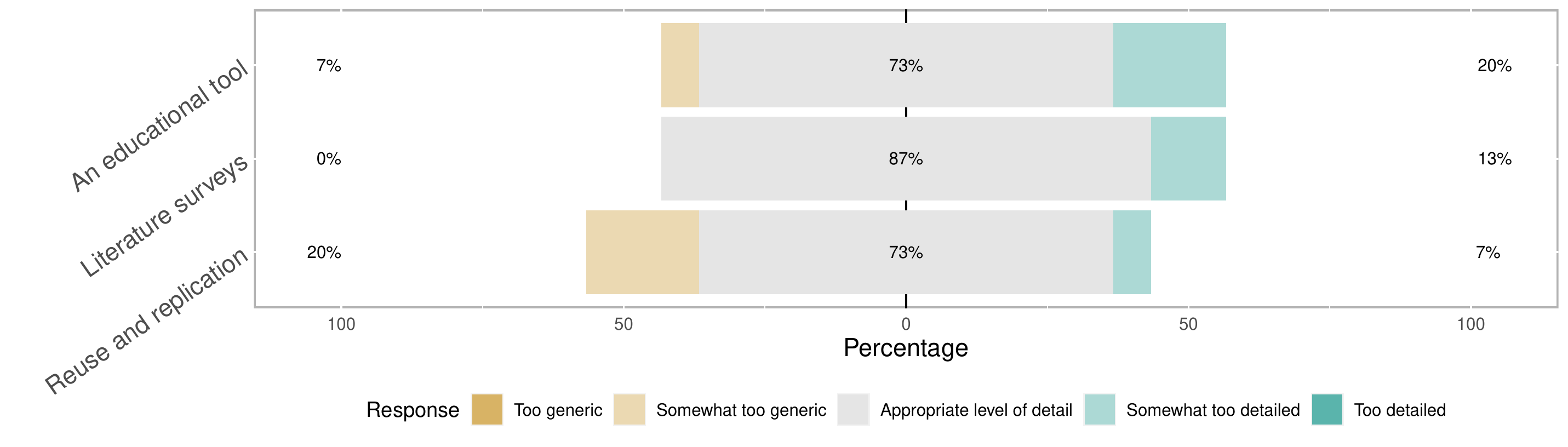}
  \caption{NLP4RE ID Card perceived level of details.}
  \label{fig:detailedness}
\end{subfigure}
\caption{Assessment of the NLP4RE ID Card appropriateness and level of details.}
\label{fig:survey}
\end{figure}

We also evaluated the respondents' perception about the \textsc{ID-card} appropriateness and level of details for the three different \rev{use cases} (see Fig.~\ref{fig:survey}).
Although the vast majority agrees that the use of the \textsc{ID-card} is appropriate across the three use cases, they perceived that more details may be necessary to further facilitate reuse and replication of existing work. 
None of the respondents indicated which details should be included, but the necessity to strike a balance between details and usability was recognised, as one of the respondent pointed out:
   \textit{``For replication purposes, I think more details would be required. On the other hand, putting in more detail will make the \textsc{ID-card} creation more time-consuming (potentially defeating the purpose). There is no easy answer here.''}
   
The respondents indicated other possible usages, such as leveraging the information contained in the card for searching and filtering previous literature, e.g., when selecting a baseline to compare in their own research work, and as a checklist to follow during the planning phase of a new work.

\section{Discussion}
\label{sec:discussion}
In the following we discuss how the \textsc{ID-card} mitigates the identified challenges, relevant hints to fill the \textsc{ID-card}, lessons learned and limitations of the study, based on the discussion carried out in the final focus group (see Step \textcircled{6} in Fig. \ref{fig:r-method}).

\subsection{Mitigation of challenges}
\rev{In Section~ \ref{sec:challenges}, we have identified several challenges concerning tool reconstruction and dataset annotation. To mitigate the above-listed challenges or reduce their effect, we suggest below some mitigation actions:
\begin{itemize}
    \item \textbf{Rigorous Annotation.} Authors should make the annotation task more rigorous by (i) conducting pilot annotations to set the protocol and to avoid later changes; (ii) maintaining written guidelines, including examples, to guide the annotators as well as for future reference; (iii) performing reconciliation sessions to resolve disagreements;  (iv) reminding the annotators about their ethical responsibilities---the annotation should be as reliable as possible, as it may be reused by other researchers in other work.
    \item \textbf{Reward reconstruction.} The research community should consider rewarding annotation and tool reconstruction tasks by creating more venues and events where such activities can be published and shared with the community.
    \item \textbf{Clarity.} The natural language description of a tool present in a paper is often insufficient and too informal for other researchers to reconstruct the tool unambiguously. Therefore, 
    researchers should allocate sufficient effort to describe how the tool can be reconstructed from scratch clearly.
    \item \textbf{Flexibility.} Flexibility is essential and often necessary to properly reconstruct existing tools. This includes simplifying or adapting existing tools to be applicable in the reconstruction scenario, e.g., using different data, filling in missing implementation details, replacing proprietary or legacy libraries.
    \item \textbf{Goal-driven Reconstruction.} The rigor of reconstruction depends on the \textit{goal} of the endeavor. For instance, building a baseline to compare a new solution requires more rigor  than for reconstructing a tool for practical use in a company. 
\end{itemize} 
}

\subsection{Guidelines for 
filling the \textsc{ID-card}}
One of the outcomes of the third focus group the authors held was the identification of a couple of considerations related to filling in the \textsc{ID-card}. In particular:
\begin{itemize}
    \item \textbf{One ID per task.} The first step when filling the \textsc{ID-card} is to map the proposed NLP solution to the underlying RE task. However, many papers discuss multiple NLP solutions (e.g., extraction and classification) in one pipeline for solving the same RE task (e.g., solving cross-referencing in requirements). 
To improve re-usability and foster reconstruction, we recommend filling the \textsc{ID-card} by focusing on one RE sub-task, which is likely the main task being solved in the paper. The \textsc{ID-card} in its current status can be theoretically filled multiple times, each time for an RE sub-task.

\item \textbf{Degree of detail.} To identify the reconstruction-relevant information, it might not be clear for the researcher filling the \textsc{ID-card} to what extent such information should be sought beyond the original paper. Although having all possible details is the best option for reconstruction, this poses a pragmatic challenge related to the additional  time required  for the respondents to fill in the \textsc{ID-card}.
The researcher should address this consideration in line with the motivation of filling in the card. \tosem{For example, the author of a paper could provide more details for better reuse and replication (see Fig.~\ref{fig:use}). 
We acknowledge that the \textsc{ID-card}s are not expected to be identical for the same paper when filled in by two different researchers, yet they would still represent equivalent summaries of the original paper. Filling in multiple \textsc{ID-card}s for the same paper is however unlikely in the future practice when the authors provide an  \textsc{ID-card} alongside their publications. 
}
\end{itemize}

\subsection{Lessons learned on the \textsc{ID-card} design}
Another outcome of the final focus group was several lessons learned during the design of the \textsc{ID-card}, which can be useful for researchers addressing similar endeavors. Among them, we remark:
\begin{itemize}
\item \textbf{Generality of the \textsc{ID-card}.} The details required by the card should cover sufficient information about the solution without merely repeating what is in the paper. The \textsc{ID-card} aims to summarize an NLP4RE paper to replicate the solution and/or dataset. Ensuring that it can cover a wide spectrum of the papers in the NLP4RE literature is important, yet challenging. There is a trade-off between the coverage of the paper by the \textsc{ID-card} versus the level of detail required to achieve a comprehensive description of the paper. Asking about many details  entails more time and effort filling in the card. 
To address this consideration, we opted to design the card at a generic-enough level to be applicable to many papers of different research focus. The \textsc{ID-card}  should help researchers decide whether or not papers are useful for their reconstruction goals.
\item \textbf{Free text options.} Missing details that are relevant to reconstruction must be properly accounted for and incorporated in the \textsc{ID-card}. We address this consideration by adding a free text option alongside each question to give the possibility to elaborate on the reason or remarks concerning missing details. We believe that providing a justification can give hints  about what to do with these missing details. For example, if the evaluation uses the entire dataset instead of cross validation, the reason might be that the tool does not require training. In this case, details about the proportion of training data might be missing but also not needed.
\end{itemize}

\subsection{Possible Uses of the \textsc{ID-card}}
\rev{The \textsc{ID-card} serves several purposes to different actors as summarized in Table \ref{tab:ID_card_uses}. To NLP4RE newcomers, e.g., researchers entering the field, PhD students starting their thesis, the \textsc{ID-card} is an effective instrument to get acquainted with the state of the art. Experienced researchers may use it when conducting their NLP4RE research at several stages, namely design, evaluation and reporting\tosem{, as a checklist to ensure that their study and covers all relevant aspects required to be replicable}. In the frequent case that the work is presented in a paper, researchers may submit the \textsc{ID-card} to a public repository as accompanying material. This way, reviewers of the paper have access to more detailed information, which helps provide an informed evaluation of the paper. Last, we foresee the \textsc{ID-card} as a useful source of information for educators when preparing NLP4RE-related material. }

We further envision that the \textsc{ID-card} can be used to create and archive summaries of papers which one has to review at a time and come back to them again at a later times---e.g., researchers performing snowballing for a literature review, can quickly look for the search seeds among such archive. 

Finally, the \textsc{ID-card} can be used to assess the adequacy for the reconstruction of a particular solution without having to read the entire paper, e.g., incomplete description, missing implementation details. In fact, it increases the chances of reconstructing a paper. In the future, \textsc{ID-card}s can be collected from the authors during the paper submission/review process and stored in a public repository to be publicly accessible by the RE community. Authors, and reviewers can also use it as a checklist of all the information that should be reported in the paper concerning reconstruction and re-usability. Thus, the \textsc{ID-card} can create more awareness of including such necessary information. 

\begin{table}[!h]
\caption{Actors using the \textsc{ID-card} as support for a number of activities. }
\label{tab:ID_card_uses}
\begin{tabularx}{0.98\textwidth} {@{} p{0.18\textwidth}p{0.22\textwidth}*{1}{>{\arraybackslash}X}@{}}
\toprule
\textbf{Actor}  & \textbf{Activity} & \textbf{Use of the \textsc{ID-card}} \\ \midrule
NLP4RE newcomer& Learn state of the art & Having a quick introduction to NLP4RE tasks using the \textsc{ID-card} structure \\ 
&  & Knowing the characteristics of existing solutions using the \textsc{ID-card} contents \\ \midrule
Study author & Design the study & Learning the relevant aspects to be covered in the study using as checklist the \textsc{ID-card} structure for the target type of NLP4RE task \\
  & Evaluate the study & Searching and filtering state-of-the-art tools that act as baseline \\
 & & Reusing or reconstructing those tools \\
& & Comparing results with respect to baseline in a systematic way \\
& Report the study & Summarizing the detailed characteristics and results of the study \\ \midrule
Paper author & Submit paper & Submitting the \textsc{ID-card} as accompanying artefact for the paper \\ \midrule
Paper reviewer & Read paper & Using the \textsc{ID-card} contents to complement the paper contents \\
& Evaluate paper & Using the \textsc{ID-card} structure as a checklist for the evaluation \\ 
& Write review & Using the \textsc{ID-card} structure to structure the review \\ \midrule
Educator & Prepare course materials & Organizing course materials using the \textsc{ID-card} structure \\
  &   & Summarizing the state of the art using the \textsc{ID-card} contents \\ \bottomrule
\end{tabularx}
\end{table}

\subsection{{Limitations}}
\label{subsec:threats}

This section identifies the limitations of our study and discusses how we mitigate them. We group the limitations distinguishing those related to the identification of the challenges, and to the construction of the \textsc{ID-card}. 

\rev{The problems of requirements ambiguity detection and requirements classification are classical and widely studied topics in NLP4RE, as shown by the survey of Zhao \textit{et al.}~\cite{zhao2021natural}. 
} However, the specific cases discussed during the focus groups are selected based on the experiences of the authors, so personal \emph{bias} might threaten the validity of the results. To mitigate personal bias, experts in different NLP4RE tasks participated in the focus groups and shared their views. 
\tosem{These focus groups involved expert NLP4RE researchers who were involved in the two presented replications; moreover, all the experts had extensive experience in various NLP4RE tasks} 
The participants also identified commonalities among the challenges identified in different focus group studies, which indicates that the challenges were not identified based on the personal bias of the participants. In order to prevent any threats due to \emph{social relations and persuasion}, each focus group was moderated by a different person, and had distinct discussion leaders. During the discussion, each participant was given the opportunity, and encouraged, to freely speak up. Nevertheless, some participants were more vocal than others. 

During the focus groups, identified challenges may not be mapped to the same \tosem{category} in the cognitive state of the participants, so \emph{understandability} of the challenges may issue a threat. To mitigate this, each moderator took notes during the focus group and shared those notes in real-time so the other participants could read and clarify them when necessary. Each moderator summarized the findings at the end of each question during the focused discussion phase and also at the end of the focus group. \emph{Completeness} of the challenges identified in the focus groups is another threat. To help achieving completeness, we did not enforce time limitations to the focus groups, allowing the participants to deeply discuss the cases and to add new points as they see fit. To ensure \emph{conciseness}, we iterated over the identified challenges to check for redundancy and reduced the redundancy. The challenges are identified based on two distinct, yet representative, NLP4RE cases. \rev{It is worth remarking that the two cases were used as triggers for the discussion, and the actual challenges elicited are based on the analysis of 46 papers, and on the substantial expertise of the participants of the focus groups, who have participated in several artifacts tracks, in NLP4RE tool design, and in study replications}. Further experimentation is however needed to ensure \emph{generality} of our findings. 
Some challenges may apply to the field of natural language processing for software engineering (NLP4SE), \rev{while others may be more general to SE.
Nevertheless, as the scope of this study is strictly NLP4RE, more studies are needed to support these claims. 
}

Two main concerns for the \textsc{ID-card} are \emph{completeness} and \emph{pragmatic usability}, which may conflict with each other. We tried to find a healthy balance of the two using multiple iterations of application and refinement of the \textsc{ID-card}\rev{, although we expect that lightweight versions of the card may be created for specific tasks and uses, e.g., for those listed in Table~\ref{tab:ID_card_uses}}. These also increased \emph{understandability} of the card as we applied it to multiple articles, and after each iteration, we discussed the results to confirm that a shared understanding was reached. Understandability was further assessed by asking the authors of the paper to fill the \textsc{ID-card}. To prevent the personal \emph{bias} of single authors affecting the construction of the \textsc{ID-card}, multiple experts of the different NLP4RE tasks participated in its construction. Issues related to \emph{coverage} of the \textsc{ID-card} in terms of types of alternative answers for each question was mitigated by assessing the card on 46 papers specifically selected to cover the spectrum of typical RE tasks, based on the taxonomy by Zhao \textit{et al.}~\cite{zhao2021natural}, and considering both widely cited seminal papers, and relevant recent ones.
\section{Conclusions and Outlook}
\label{sec:conclusion}

Replication, covering both data (re-)annotation and tool reconstruction, is an important strategy in experimentation and empirical evaluation. 
In this paper, \rev{addressing the field of NLP4RE, }
  we \tosem{investigated what are the challenges of annotating datasets for training and evaluating NLP4RE tools (RQ1), and the challenges in reconstructing NLP4RE tools (RQ2). To answer these research questions, we conducted focus groups where we reflected on our first-hand experience in replicating NLP4RE state-of-the-art tools, and we further analyzed 46 papers covering a wide spectrum of NLP4RE landscape. As a result of our study, we identified 10 challenges concerning data annotation and 5 challenges for reconstructing tools. Some of these challenges are specific to  NLP4RE, while part of them can be considered applicable to other software engineering fields. Though we refrain from generalising our findings as they stem from an NLP4RE context, we encourage other authors to further investigate our list of challenges, and possibly adapt it to their contexts.}
 
\tosem{Challenges concerning data annotation (RQ1) include: the unavailability of theories specific to RE  tasks to build the annotation task on, e.g., concrete definitions of non-functional requirements or nocuous ambiguity; the need to anticipate additional time and effort to potentially evolve annotation protocols; and issues resulting from dataset imbalance and the lack of domain knowledge of annotators. 
Challenges for reconstructing tools (RQ2) mostly arise from missing details in the original papers, e.g., the exact library version, or the application of outdated NLP tools that are no longer available.
 }

To reduce the effect of these challenges, \tosem{we investigated how to support NLP4RE researchers (RQ3). } 
We proposed an \textsc{ID-card} as a complementary source to original papers for summarizing\tosem{, via a total of \rev{47} questions, information relevant to replication}. The \textsc{ID-card} covers seven topics \tosem{essential for replication. These concern the } RE and NLP tasks, inputs, outputs, annotation process, tool, and evaluation details. 
We assessed the \textsc{ID-card} both internally and externally (with the authors of original papers) by \tosem{cross-filling of } 
the same paper. Though filling the \textsc{ID-card} requires time and effort (which should be marginal for the authors of a paper), the \textsc{ID-card} provides a useful starting point to facilitate replication.  

In the future, we would like to foster the \textsc{ID-card} as part of paper submission at different RE venues. The goal is to help reviewers assess the replication consideration of the paper and increase the overall awareness of replication-relevant information. 
Furthermore, we plan to investigate extending the \textsc{ID-card} to cover other software engineering-related areas. 
This is highly needed as artifact evaluation is not sufficiently mature, and needs further improvement~\cite{hermann2020community,winter2022retrospective}.
Though the \textsc{ID-card} mainly aims to improve artifact evaluation, this is not its only purpose, as it can be used for other objectives, such as literature reviews, as also remarked by the participants in our external assessment. These are hypothetical scenarios of usage, which can evolve after the \textsc{ID-card} is used and possibly adapted by the community. 

%

\section*{Acknowledgments}
This paper was partially supported by the following projects and grants: Italian MUR--PRIN 2020TL3X8X project T-LADIES (Typeful Language Adaptation for Dynamic, Interacting and Evolving Systems); 
EU Project CODECS GA 101060179, by the MOST – Sustainable Mobility National Research Center and received funding from the European Union Next-Generation EU (PIANO NAZIONALE DI RIPRESA E RESILIENZA (PNRR) – MISSIONE 4 COMPONENTE 2, INVESTIMENTO 1.4 – D.D. 1033 17/06/2022, CN00000023); 
KKS foundation through the S.E.R.T. Research Profile project at Blekinge Institute of Technology; 
Spanish Ministerio de Ciencia e Innovación under project /funding scheme PID2020-117191RB-I00/AEI/10.13039/501100011033.


\bibliographystyle{ACM-Reference-Format}
\bibliography{main}

\appendix
\section{Details of the Focus Groups}\label{sec:fg}

\begin{table}[htbp]
\centering
\caption{\rev{Moderators, leads, and participants of the focus groups. }}
\label{tab:mods}
\rev{
\begin{tabular}{@{}llll@{}}
\toprule
    & Moderator & Leads          & Participants   \\ \midrule
Focus Group RQ1 & A2        & A1, A3, A4, A6 & A5, A7         \\
Focus Group RQ2 & A5        & A1, A2, A4,    & A3, A6, A7     \\
Focus Group ID Card & A6        & A1, A3         & A2, A4, A5, A7 \\ \bottomrule
\end{tabular}%
}
\end{table}

\sectopic{Protocol. }
Table~\ref{tab:mods} reports the moderators, lead participants, and regular participants for the three focus groups. All authors are experienced in the annotation and tool reconstruction tasks. 
We follow the guidelines by Breen~\cite{breen2006practical}: 
\begin{description}
    \item[\textbf{Focus Group Preparation.}] This task includes the following steps:
     \begin{enumerate}
        \item A moderator is appointed among the \rev{authors who did not participate in the annotation (focus group 1) or reconstruction cases (focus group 2), or ID card definition (focus group 3)}.
        \item \rev{Lead participants are selected among those who participated in the annotation, reconstruction, and ID card definition cases. Regular participants are those who are not lead participants or moderators.  }
        \item A meeting is organised in which the moderator is informed about the results by the \rev{lead participants}, and is given an overview of the collected data.
        \item The moderator defines a schedule for the focus group and organises it.
        \item Before each focus group, the lead participants are required to independently prepare a list of challenges or relevant discussion topics about the experience. These are shared during the focus group execution.
    \end{enumerate}     
    \item[\textbf{Focus Group Execution.}] 
The focus group session is planned for 90 minutes: 40 minutes for free discussion, 40 minutes for focused discussion, and finally 10 minutes for wrapping up the session.
The moderator starts the session by introducing the main question to be addressed by the focus group, i.e., the RQ associated to the focus group. 
Following this, the lead participants present their experience to the others in two slides prepared in advance prior to the actual session. The moderator then collects a list of challenges mentioned by the lead participants in one slide and moderated the discussion among the participants in free-form manner.
Further questions are progressively added and addressed within eight minutes each during the focused discussion. After the focused discussion, the moderator summarizes the discussion and concludes the session. 
    \item[\textbf{Qualitative Data Collection and Analysis.}] The moderator of the focus group analyses the tape recording, and extracts \rev{the themes that were discussed the most and perceived as the most important}. \rev{This was done by the moderator to ensure neutrality, as the moderator was someone who did not directly participated in the activity discussed in each focus group}.
    \item[\textbf{Review of Qualitative Data.}] The themes are proposed to the participants in a wrap-up meeting. \rev{This is conducted two weeks after each focus group, lasts 9 minutes, and involves all participants.} The participants can complete, enrich or amend the themes to come to an agreed set of challenges around the specific RQ.
\end{description}

\sectopic{Dataset Annotation and Re-annotation Focus Group Details }
This first focus group aimed to discuss the challenges related to data annotation for NLP4RE tasks as described in Sec.~\ref{sec:rm}.
It started with the presentations of the lead participants for the free discussion. During the free discussion, 18 challenges were listed by the participants and documented by the moderator in a shared document. The free discussion was followed by the questions asked by the moderator for the focused discussion. Below, we list these questions and summarize the discussion items.

\begin{itemize}
    \item \textbf{Q:} How do you decide on the labels for the data?
    \textbf{A:} Several factors affect the decision on the labels: \emph{(i)} existing theory or \emph{(ii)} the limitations of the state-of-the-art can be the sources for the set of labels. At the same time, \emph{(iii)} the recency of the labels used in the state of the art is a determining factor on the decision whether or not to update the set of labels. Another factor is \emph{(iv)} the availability of the raw data: following an opportunistic approach, the researchers may add or drop some labels. Finally, \emph{(v)} assumptions from linguistic and industrial background may have an impact on the set of labels (e.g., linguistics ambiguity does not necessarily lead to multiple interpretations in software development due to company conventions and standards).
    \item \textbf{Q:} How do you construct the annotation protocol?
    \textbf{A:} Protocol construction is an iterative process. An initial (pilot) annotation helps to construct a protocol. The annotators need to seek a balance between the effort put in the annotation process and the cost of validation and re-annotation resulting from potential changes in the protocol. The participants also observed that the protocols do not guarantee an agreement among the annotators, and even when the protocols are documented as written guidelines, human aspects such as fatigue handling or persuasion avoidance are typically not documented in the annotation protocol. \rev{This answer summarizes the experience of the authors on annotating data for NLP4RE tasks.}
    
    \item \textbf{Q:} How do you validate the labels and annotation protocol?
    \textbf{A:} The validation is done by checking the labels, and the protocols are aligned with the purpose of the annotation process. Due to practical considerations, researchers typically proceed with a good-enough protocol. \rev{The consensus is that the annotators do not have strict protocols as in medical studies. Annotation protocols are more flexible and prone to change during the annotation process. Since trials and changes are costly, the researchers are contempt with good-enough nannotation protocols and they do not strive for the perfect ones.}
    
    \item \textbf{Q:} What are the requirements for the annotators?
    \textbf{A:} Independence of the annotators is a desired quality, which is difficult to satisfy unless there is a budget for the annotators. For some annotation tasks, domain knowledge can be crucial to ensure reliable outcome. 
     Experience in annotation, whether for requirements engineering or not, is also desired. When seeking skilled annotators, crowd-sourcing platforms may output annotations with a questionable quality. Doctoral students, on the other hand, can efficiently provide good quality results. 
    
    \item \textbf{Q:} How are the annotators trained?
    \textbf{A:} Written guidelines are used as the training material. Pilot sessions and annotating in small iterations also contribute to the training process. Yet there is no formal training on the group theory\rev{~\cite{johnson1991joining}} and collaborative decision making and those skills are acquired through experience. 
    
    \item \textbf{Q:} What are the format and tools used for annotation?
    \textbf{A:} Practical concerns determine the tool for the annotation process and the format of the annotated data. Simple tools such as spreadsheets and common \rev{file} formats such as comma-separated values (CSV) are preferred. 
    
    \item \textbf{Q:} When do you stop annotation?
    \textbf{A:} The size of the annotated data, class balance\rev{---having a comparable number of samples for each class---}, and practical concerns such as the availability of the annotators can be important factors for this decision which is taken during the process. 
    
    \item \textbf{Q:} Is annotation a rewarding experience?
    \textbf{A:} The focus group participants found the annotation process both rewarding and unrewarding. While the annotation process itself can be interesting and triggering for some, providing a deeper understanding on the problem and on research as well, it may quickly become tedious and not challenging, especially with the increase of the size of the dataset. Nevertheless, the participants found that sharing the dataset could be rewarding as it is used by the community.  
    
\end{itemize}

\sectopic{Tool Reconstruction Focus Group Details.}
This focus group aimed to discuss and reflect on our first-hand experience in reconstructing the two state-of-the-art solutions introduced in Section~\ref{sec:cases}. 

The moderator started the session by introducing the main question to be addressed by this focus group, which is ``What are the challenges in reconstructing NLP4RE tools?''. Following this, the lead participants presented their experience to the others in two slides prepared in advance prior to the actual session. The moderator then collected a list of 23 challenges mentioned by the lead participants in one slide and moderated the discussion among the participants in free-form manner. The study then continued with the focused discussion where further questions were progressively added and addressed within eight minutes each during the focused discussion. Below, we list the questions and summarize the discussion items.
\begin{itemize}
    \item \textbf{Q:} Where do you start when reconstructing tools?
    \textbf{A:} For reconstructing a tool, one starts by (1) defining the goals (e.g., benchmark or baseline), (2) selecting a state-of-the-art paper that is clear enough, (3) identifying the tool components to be reconstructed, and (4) adapting the tool to fit the available dataset. 

    \item \textbf{Q:} Do you follow a structured and repeatable process for tool reconstruction?
    \textbf{A:} Reconstruction is often an ad-hoc process depending on the underlying case, reconstruction goals and what is publicly available (e.g., source code). 

    \item \textbf{Q:} How do you ensure that the reconstruction is correct?
    \textbf{A:} Unless the original tool is publicly available, comparison is not always possible due to lack of implementation details (e.g., hyper-parameters or library versions). However, one should use the same evaluation metrics, and validate intermediate steps using simple examples. 
 
    \item \textbf{Q:} What are the most time-consuming phases of tool reconstruction?
    \textbf{A:} Adapting the original tool to fit the reconstruction goals requires deriving the implementation details from the original paper. Most often, time and effort are put into checking alternatives, and making decisions about the unknowns (e.g., which parameter value, imbalance handling, thresholds).
    
    \item \textbf{Q:} Do you consider tool reconstruction a rewarding experience? (Would you do it again?)
    \textbf{A:} The participants agreed that reconstruction is challenging. At the individual level, it could be to some extent rewarding as one gets the opportunity to learn about the process, spot incorrectness or conceptual errors. At a community level, reconstruction can be rewarding if the reconstructed tool is shared, e.g., as a baseline. 
\end{itemize}

Finally, additional issues emerged during the focus group. In summary, dealing with the unknowns can be an important factor for determining the divergence level between the reconstructed tool and the original one. Interaction with the authors is another factor that can facilitate the reconstruction process; however, since maintaining a tool is not common within the academic research community, authors' help might be limited.
\newline

\sectopic{ID Card Design Focus Group Details}  
The focus group session followed a similar structure like the previous ones. More concretely, the seven researchers (co-authors of this paper) participated in this focus group as one moderator, two lead participants, and four discussants.

In this case, the focus group was moderated by one of the authors \rev{(A6)}  who did not participate in the development of the ID card. The moderator started the session by introducing the main question to be addressed by this focus group, which is ``What are the challenges in extracting replication-relevant information from NLP4RE papers?''. Following this, the lead participants presented their experience to the others in two slides prepared in advance prior the actual session. The moderator then collected a list of 21 challenges mentioned by the lead participants in one slide and moderated the discussion among the participants in free-form manner. The challenges were grouped into five categories, defining content, designing questions and answers, evaluation and interrater agreement, paper-structure related and other. 
Further questions were progressively added and addressed within eight minutes each during the focused discussion. Below, we list the questions and summarise the discussion items.

\begin{itemize}
    \item \textbf{Q:} Is there any aspect that the ID card does not reflect? 
    \textbf{A:} Some aspects were intentionally left out, e.g., some RE tasks, or details concerning manual steps which involve interaction with users. This choice was motivated by three reasons. First, the ID card should be generic enough to cover a wide spectrum of RE literature, and thus instead of specifying all details, we substituted the omitted choices by open questions that allow free text answers. Second, the ID card should not be regarded as a replacement of a paper, rather a complementary resource that provides structured summary. For example, context that is part of the paper should not necessarily be part of the ID card. Finally, limiting the ID card with a finite list of answers (choices) reduces the usability of the ID card in other contexts, e.g., software engineering tasks that are not part of the NLP4RE literature.  

    \item \textbf{Q:} Is it necessary to complement the provided answers with explanations, justifications, or any evidence?
    \textbf{A:} Though explanations are not enforced in the ID card, they are still encouraged. We have added a free-text choice to each question to provide the tagger with the possibility to complement a given answer with further details, justifications or evidence.  
    
    \item \textbf{Q:} How trustworthy is the collected information? How do annotators’ disagreements affect the overall quality of the evaluation? 
    \textbf{A:} We observed that most disagreements are about the level of details that the annotators provided rather than actual contradictions. 
    Thus, we believe that the ID card is trustworthy as a companion to the paper that facilitates the access to 
    replication information, the original purpose for which it is designed. 

    \item \textbf{Q:} How can the ID card be used in the context of replication? 
    \textbf{A:} Since the ID card provides a structured summary of the original paper, it informs the researcher about the suitability of the study for their replication scenario. To increase the benefit of the ID card, we envisage appending it as supplementary material to future papers and/or collect the ID cards in one public repository to be shared with the community. The ID card can further provide guidelines for reviewing papers. 
    Generalizing the ID from NLP4RE to other (more general) domains, e.g., NLP4SE, might require effort but is rewarding in the long term. 
    
    \item \textbf{Q:} Do you consider filling the ID card a rewarding experience? (Would you do it again?) \textbf{A:} \rev{According to the participants in the focus group, filling in the ID card would
    require less effort and time from the author of the paper\footnote{\rev{During the focus group, the participants have agreed on the hypothesis that filling the ID card for a paper would require less time and effort if done by the author of that paper. Considering the responses on the ease of use of the ID card (see Fig. 3 in Section 5 of our paper), we observed that this hypothesis is not always true in particular when the author fills the ID card for a paper that has long been published.}}. Thus, the participants agree on the advantage of encouraging the creation of an ID card to be publicly released alongside the paper.}
\end{itemize}

\end{document}